\pdfoutput=1

\documentclass[11pt,twoside,a4paper,cmspaper,final,collab]{cms-tdr}
\def\svnVersion{73010:73012}\def\cmsCernNo{2011-093}\def\cmsCernDate{\today}\def\cmsMessage{Submitted to the Journal of High Energy Physics}

\begin{document}\cmsNoteHeader{EWK-10-007}

\hyphenation{had-ron-i-za-tion}
\hyphenation{cal-or-i-me-ter}
\hyphenation{de-vices}
\RCS$Revision: 73012 $
\RCS$HeadURL: svn+ssh://alverson@svn.cern.ch/reps/tdr2/papers/EWK-10-007/tags/arxiv-0/EWK-10-007.tex $
\RCS$Id: EWK-10-007.tex 73012 2011-08-02 11:12:29Z alverson $
\hyphenation{mar-gin-al-ly}
\newcommand{\pp}{\ensuremath{\mathrm{pp}}}%
\newcommand{\ppbar}{\ensuremath{\mathrm{p}\bar{\mathrm{p}}}}%
\newcommand{\ee}{\ensuremath{\mathrm{e}^{+}\mathrm{e}^{-}}}%
\renewcommand{\ttbar}{\ensuremath{\mathrm{t}\bar{\mathrm{t}}}}%

\cmsNoteHeader{EWK-10-007} 
\title{Measurement of the Drell--Yan Cross Section  in pp Collisions at $\sqrt{s} = 7\TeV$}

\date{\today}

\abstract{
  The Drell--Yan differential cross section is measured in
$\ensuremath{\mathrm{pp}}$
  collisions at
  $\sqrt{s} = 7$ TeV,
  from a data sample
  collected with the CMS detector at the LHC, corresponding to an integrated luminosity of $36~{\mathrm{pb}}^{-1}$.
  The cross section measurement, normalized to the measured cross section in the Z region, is reported for both the dimuon and dielectron channels in the
dilepton invariant mass range
15--600 GeV.
The normalized cross section values are quoted both
in the full phase space and within the detector acceptance. The effect of final state
radiation is also identified. The results are found to agree with theoretical predictions.

}

\hypersetup{%
pdfauthor={CMS Collaboration},%
pdftitle={Measurement of the Drell-Yan Cross Section in pp Collisions at sqrt(s) = 7 TeV },%
pdfsubject={CMS},%
pdfkeywords={CMS, physics, EWK, Drell Yan}}

\maketitle 

\newcommand{\DYee}{DY\ensuremath{\;\rightarrow \mathrm{e}^+\mathrm{e}^-}}    
\newcommand{\DYmumu}{DY\ensuremath{\;\rightarrow \mu^+\mu^-}}    
\newcommand{\DYtt}{DY\ensuremath{\;\rightarrow \tau^+\tau^-}}    
\newcommand{\NGEN}{N_{\mathrm{gen}}}
\newcommand{\NACC}{N_{\mathrm{acc}}}
\newcommand{\NEFF}{N_{\epsilon}}
\newcommand{\MLL}{M(\ell\ell)}

\newcommand{\rhoeff}{\rho_{\mathrm{eff}}}
\newcommand{\effdata}{\varepsilon_{\mathrm{data}}}
\newcommand{\effmc}{\varepsilon_{\mathrm{sim}}}

\newcommand{\effreco}{\varepsilon_{\mathrm{reco}}}
\newcommand{\effid}{\varepsilon_{\mathrm{id}}}
\newcommand{\efftrig}{\varepsilon_{\mathrm{trig}}}

\newcommand{\ANORM}{A_{\mathrm{norm}}}
\newcommand{\LUM}{{\cal{L}}}
\newcommand{\NUNORM}{N_{u,{\mathrm{norm}}}}
\newcommand{\effNORM}{\varepsilon_{\mathrm{norm}}}
\newcommand{\rhoNORM}{\rho_{\mathrm{norm}}}

\newcommand{\RPOSTFSR}{R_{\text{post-FSR}}}
\newcommand{\RPOSTFSRDET}{R_{\text{det,post-FSR}}}
\newcommand{\RDET}{R_{\mathrm{det}}}

\def\dsdm{\ensuremath{{\mathrm d}\sigma/{\mathrm d}M}}

\section{Introduction}
\label{sec:introduction}
 
The production of lepton pairs in hadron-hadron collisions via the Drell--Yan (DY) process
is described in the standard model (SM) by the $s$-channel exchange of $\gamma^*/\text{Z}$.  
Theoretical calculations of the differential cross section $d\sigma/d\MLL$, where $\MLL$ is the 
dilepton invariant mass, are well established up to 
next-to-next-to-leading order (NNLO)~\cite{DY-theory, DYNNLO, DYNNLO1}.
Therefore, comparisons between calculations and precise experimental measurements
provide stringent tests of perturbative quantum chromodynamics (QCD) and significant 
constraints on the evaluation of the parton distribution functions (PDFs).
Furthermore, the production of DY lepton pairs constitutes a major source of background
for \ttbar~ and diboson measurements, as well as for searches for new physics, such as 
production of high mass dilepton resonances.

This paper presents a measurement of the differential DY cross section
in proton-proton collisions at $\sqrt{s} = 7\TeV$, based on dimuon and dielectron 
data samples collected in 2010 by the Compact Muon Solenoid (CMS) experiment at the Large
Hadron Collider (LHC), corresponding to an integrated luminosity of $35.9\pm 1.4$\pbinv.
The results are given for the dilepton invariant mass range $15 < \MLL < 600\GeV$, corresponding to the 
Bjorken $x$ range 0.0003--0.633 for the interacting partons, and complement the observations 
previously reported by the Tevatron collaborations~\cite{D0,CDF_a,CDF_b}.
To reduce systematic uncertainties, the results are normalized to the cross section in the
$Z$~region ($60 < \MLL < 120\GeV$) 
as determined in the same measurement.
The inclusive Z cross section in the full phase space was measured previously by
CMS~\cite{ZCrossSection}. 

In the analysis presented, the cross sections are calculated as 

\begin{equation}\label{eqn:fullCrossSection_intro}
\sigma = \frac{N_{\text{u}}}{A \, \epsilon \, \rho \, \LUM}\, , 
\end{equation}

where $N_{\text {u}}$ is the unfolded background-subtracted yield, corrected for detector 
resolution.  The values of the acceptance $A$ and the efficiency $\epsilon$ are 
estimated from simulation, while $\rho$ is a factor that accounts for differences 
in the detection efficiency between data and simulation.  
Knowledge of the integrated luminosity $\LUM$ is
not required for the measurements described in this paper, since the cross
sections are normalized to the Z region.

This paper is organized as follows: in Section \ref{sec:CMS-detector} the CMS detector is described, with particular 
attention to the subdetectors used to identify charged leptons.
Section \ref{sec:Event-Selection} describes 
the data and Monte Carlo (MC) samples
used in the analysis and the selection applied to identify the DY candidates.
The signal extraction methods for the muon and electron channels, 
as well as the background contributions to the candidate samples are discussed in Section
\ref{sec:backgrounds}. Section \ref{sec:unfolding} describes the analysis techniques used to unfold the detector resolution 
from the measurements.
The calculation of the geometrical and kinematic acceptances together with 
the methods applied to determine
the reconstruction, selection, and trigger efficiencies of the leptons within the experimental acceptance are presented in Section \ref{sec:AccepEff}. 
Systematic uncertainties are discussed in Section \ref{sec:systematics}. 
The calculation of the shapes of the DY invariant mass distributions 
are summarized in
Section \ref{sec:results}. In that section we report not only results in 
the full phase space but also results
as measured within the fiducial
and kinematic acceptance (both before and after final state QED radiation 
corrections), thereby eliminating the PDF uncertainties from the results.

\section{The CMS Detector}
\label{sec:CMS-detector}

A detailed description of the CMS detector and its performance can be found
in Ref.~\cite{ref:CMS}. The central feature of the CMS apparatus is a
superconducting solenoid 13~m in length and 6~m in diameter, which
provides an axial magnetic field of 3.8~T. Within the field volume are
the silicon pixel and strip tracker, the crystal electromagnetic
calorimeter (ECAL), and the brass/scintillator hadron calorimeter. 
Charged particle trajectories are measured by the
tracker, covering 
 the full azimuthal angle
and pseudorapidity interval $|\eta| < 2.5$, 
where the pseudorapidity is defined as 
$\eta = -\ln \tan (\theta/2)$,
with $\theta$ being the polar angle of the trajectory of the particle
with respect to the counterclockwise beam direction.  
Muons are measured in the pseudorapidity range $|\eta|< 2.4$, 
with detection planes made using three technologies: drift tubes, cathode 
strip chambers, and resistive plate chambers. The muons associated with the 
tracks measured in the silicon tracker have a transverse momentum (\pt)
resolution of about 2\% in the muon \pt range relevant for the analysis 
presented in this paper.
The ECAL consists of nearly 76\,000 lead tungstate
crystals, distributed in a barrel region ($|\eta| < 1.479$) and
two endcap regions ($1.479 < |\eta| < 3$), and 
has an ultimate energy resolution better than 0.5\%
for unconverted photons with transverse energies (\ET) above 100~\GeV. The
electron energy resolution is better than 3\% for the range of
energies relevant for the measurement reported in this paper. 
A two-level trigger system selects the 
events for use in 
offline physics analysis.

\section{Event Selection}
\label{sec:Event-Selection}

The basic signature of the DY process is straightforward: 
two oppositely charged isolated leptons originating in the same primary vertex. 
The analysis presented in this paper is based on dilepton data samples
selected by inclusive single-lepton triggers. 
The dimuon data sample was selected by a single-muon trigger with a \pt 
threshold ranging from 9 to 15\GeV, depending on the beam conditions. 
In the offline selection, one of the muons is required to match, 
in three-dimensional momentum space, a muon trigger candidate, and must have 
$|\eta| < 2.1$ and $\pt > 16$\GeV, to ensure that it is on the plateau of the trigger 
efficiency curve. The second muon is required to have $|\eta| < 2.4$ and $\pt > 7$\GeV.
No muon isolation is required at the trigger level.

Muons are required to pass the standard CMS muon
identification and quality criteria, based on the number of hits found
in the tracker, the response of the muon chambers, and a set of
matching criteria between the muon track parameters as determined by
the inner tracker section of the detector and as measured in the muon 
chambers~\cite{tag-and-probe, ref:CRAFT}. 
These criteria ensure 
that only muons with well-measured parameters are selected for the 
analysis. To eliminate cosmic-ray muons, each muon is 
required to have an impact parameter in the transverse plane 
less than 2~mm with respect
to the center of the interaction region, and the opening 
angle between the two muons must differ from $\pi$ by more than 5~mrad.
In order to reduce the fraction of muon pairs from (different) light-meson decays 
a common vertex for the two muons is fitted 
and the event is rejected if the
dimuon vertex $\chi^2$ probability is smaller than 2\%. 
Finally, an isolation requirement is imposed on both muons,
$I_{\mathrm{rel}} = (\sum \pt({\mathrm{tracks}}) + \sum \ET({\mathrm{had}})) /\pt(\mu) < 0.15,$
where $\sum \pt({\mathrm{tracks}})$ is the sum of the transverse 
momenta of all the additional tracker tracks and $\sum \ET({\mathrm{had}})$ is the sum of all 
transverse energies of hadronic deposits in a cone 
$\Delta R = \sqrt{(\Delta\phi)^2+(\Delta\eta)^2} < 0.3$
centered on the muon direction and excluding the muon itself. 
Given that muons can radiate nearly collinear photons
in a process referred to as final state electromagnetic radiation (FSR),
deposits in the ECAL are not included in the definition.
Otherwise an inefficiency would be 
introduced in the analysis.

For the electron analysis, events are selected with a trigger requiring at least one electron,
with a minimum \ET ranging from 15 to 17~\GeV, 
depending on the 
beam conditions.
Electron reconstruction starts from clusters of energy deposited in
the ECAL, and associates with them hits
in the CMS 
tracker~\cite{EGMPAS}. 
Energy-scale corrections are
applied to individual electrons as described in Ref.~\cite{WAsymmetry}. 
The electron candidate is required to be consistent with 
a particle originating from the primary vertex in the event. Electron 
identification criteria based on shower shape and track-cluster 
matching are applied to the reconstructed candidates.
Electrons originating from photon conversions are rejected by 
eliminating those electrons for which a partner track consistent with
a conversion hypothesis is found, and requiring no missing hits
in the pixel detector,
as discussed in Ref.~\cite{ZCrossSection}.
Isolation requirements are imposed on each electron, according to 
$(\sum \pt({\mathrm{tracks}}) + \sum \ET({\mathrm{had}}) + \sum \ET({\mathrm{em}}) ) /\pt({\mathrm{e}}) < 0.1,$
where  $\sum \pt({\mathrm{tracks}})$ and $\sum \ET({\mathrm{had}})$ are 
defined as explained for muons,
and $\sum \ET({\mathrm{em}})$ is the sum of the transverse energies of electromagnetic deposits
in $\Delta R < 0.3$, excluding the electron candidate itself.
The standard CMS isolation calculation for electrons also excludes 
ECAL energy deposits that are potentially created by FSR photons, while
absorbing some of these deposits into electron objects. Thus,
the FSR-related inefficiencies, present for muons, are avoided for
electrons and ECAL information is used in the total isolation calculation.
The criteria were optimized to maximize the rejection of misidentified
electrons from QCD multijet production and nonisolated electrons
from heavy-quark decays, while maintaining at least 80\% efficiency
for electrons from the DY process. More details are 
found in Ref.~\cite{ZCrossSection}.

Electrons must be reconstructed in the ECAL barrel
with $|\eta| < 1.44$ or in the ECAL endcaps with
$1.57 < |\eta| < 2.5$.
The leading electron is required to have $\ET > 20$\GeV, 
while the second electron must have $\ET > 10$\GeV.
The leading electron in a candidate pair is required
to match, in $\eta$ and $\phi$, a trigger electron candidate.

Event samples for simulation studies of electroweak
processes involving W and Z production are produced with the NLO 
MC
generator
{\sc POWHEG}~\cite{Alioli:2008gx, Nason:2004rx, Frixione:2007vw} interfaced
with the {\sc PYTHIA} (v.~6.422)~\cite{Sjostrand:2006za} parton-shower event generator, using
the CT10~\cite{CT10} parametrization of the PDFs. {\sc PYTHIA} is also used for the FSR simulation. 
The QCD multijet background is
generated with {\sc PYTHIA}, and  
the \ttbar~ background is simulated using {\sc MadGraph} (v.~4.4.12)~\cite{MadGraph} and {\sc PYTHIA}
at leading order using the CTEQ\,6L PDF set~\cite{CTEQ} for both samples. 
Generated events are processed through the full {\sc GEANT4}~\cite{GEANT4}
detector simulation, trigger emulation, and event reconstruction chain.

\begin{figure}[t!]
  \begin{center}
    \includegraphics[width=0.47\textwidth, angle=90]{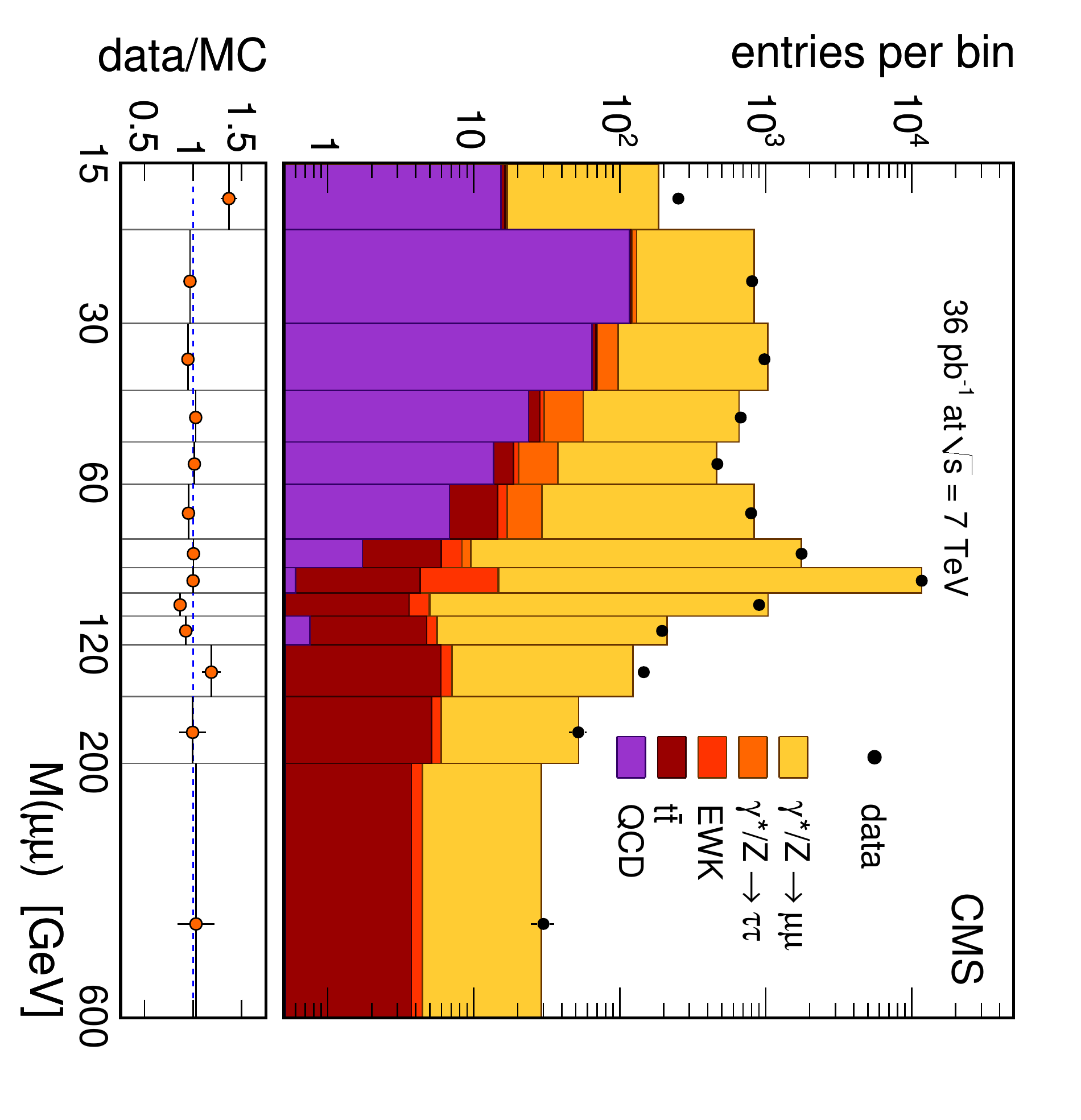}
    \includegraphics[width=0.47\textwidth, angle=90]{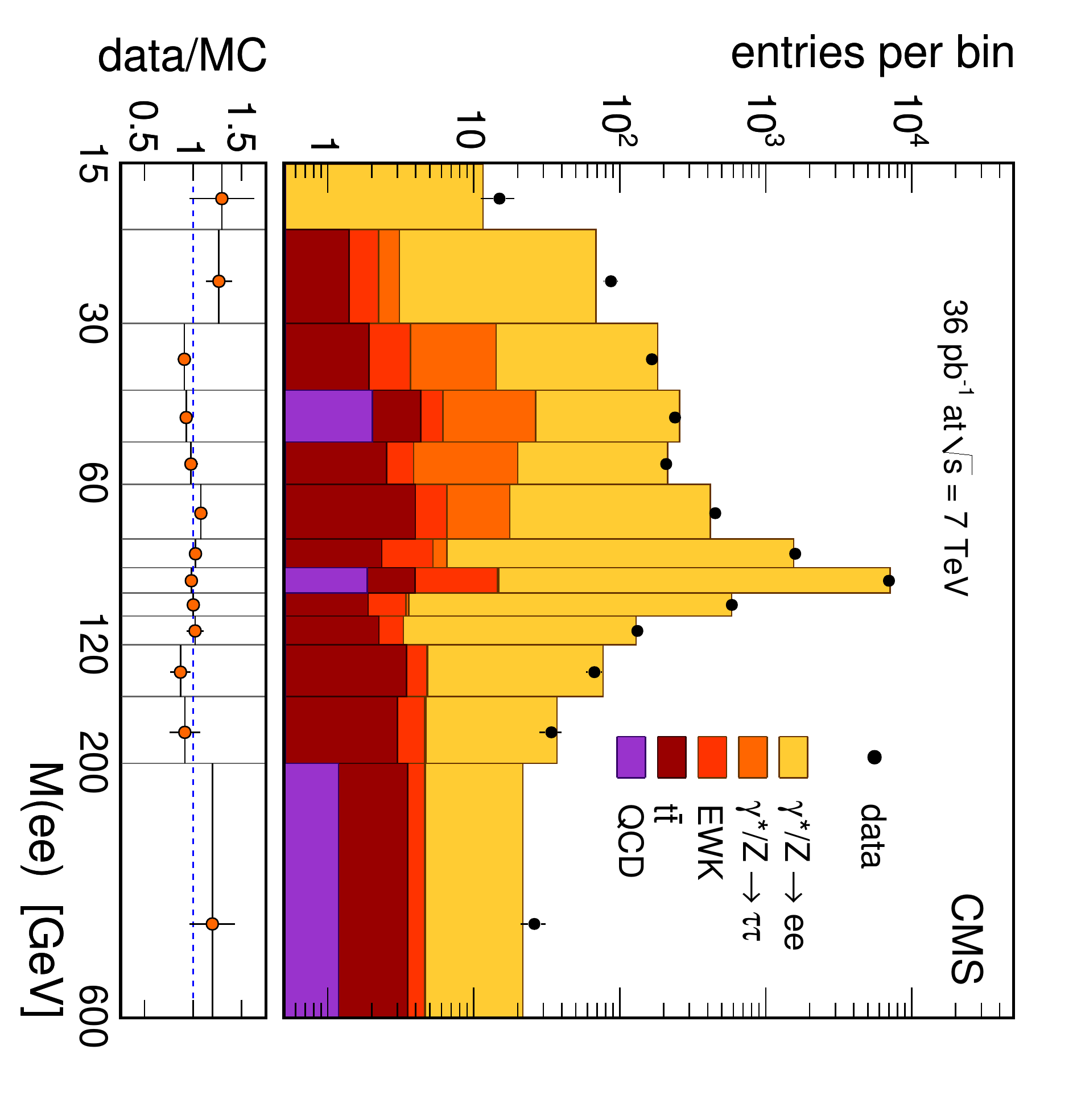}
    \caption{The observed dimuon (left) and dielectron (right) invariant mass spectra. No corrections are applied to the distributions. The points with error bars represent the data, while the 
various contributions from simulated events are shown as
stacked histograms. By ``EWK'' we denote $Z/\gamma^* \rightarrow \tau \tau$, $\text{W} \rightarrow \ell\nu$, and diboson production. The ``QCD'' 
contribution results from processes associated with QCD and could be genuine or misidentified leptons.  
The lower panels show the ratios between the measured and the 
simulated distributions including the statistical uncertainties from both.
       \label{fig:mass-observed}
    }
  \end{center}
\end{figure}

The observed invariant mass distributions, in the dimuon and dielectron channels, are shown in 
Fig.~\ref{fig:mass-observed}. Thirteen mass bins are used to cover the observable dilepton mass
spectrum. These are chosen not only to be wide enough to minimize the influence of the mass resolution but 
also to provide good statistical power. The mass resolution varies between a few hundred MeV at the
low invariant masses covered and several tens of GeV at the high end of the spectrum. The mass 
bins have unequal widths.             

\section{Backgrounds}
\label{sec:backgrounds}

Several physical and instrumental backgrounds contribute to both the 
dimuon and dielectron analyses.
The main backgrounds at 
high dilepton invariant masses are caused by \ttbar\ and diboson production, while at  
invariant masses below the $Z$ peak, DY production of $\tau^+\tau^-$ pairs 
becomes the dominant background. At low dimuon invariant masses, most background events 
are QCD multijet events.
The expected shapes and 
relative yields of these several dilepton sources can be seen in 
Fig.~\ref{fig:mass-observed}.

For the dimuon channel, the electroweak and \ttbar\ backgrounds are evaluated through
simulation studies, expected to provide a good description of the real contributions.
This is also verified by related studies in the electron channel presented below.
In contrast, the QCD background is evaluated from data by two independent  
methods. 
The first estimates the yield of opposite-sign (OS) background muon pairs by scaling 
the yield of same-sign (SS) pairs. The scaling is based on information from the ratio of 
OS/SS events when one of the muons is not isolated (a sample dominated by background),
and the MC prediction that the same ratio holds when both muons are isolated.
Statistical uncertainties in all the cases are propagated to the final background 
estimate.
The second method, which is more precise, is based on 
the signal/background discriminating variable $I_{\mathrm{rel}}$.
We obtain $\pt$-dependent isolation distributions (templates) from almost pure samples of background 
and signal events, respectively composed of SS and OS muon pairs. The latter consist of events in 
the Z mass peak surviving tight quality selection criteria.  A superposition of these two shape distributions 
is fitted to the observed isolation distributions of the two muons, for each invariant mass bin. The dimuon invariant mass 
distribution of the QCD background is obtained as the weighted average of the estimates from the two methods.

There are two categories of dielectron backgrounds: the first category 
contributes candidates composed of two genuine electrons and the second
contributes candidates 
in which at least one particle
is a misidentified electron.
Most of the genuine dielectron background is due to \ttbar, WW, and tW
production, as well as DY production of $\tau^+\tau^-$ pairs. 
We estimate the contribution from these processes with a sample of 
${\mathrm{e}}^{\pm}\mu^{\mp}$  events having the same
physical origin.
This signal-free sample contains approximately twice the estimated number of 
background events contaminating the ${\mathrm{e}}^+{\mathrm{e}}^-$ sample, and provides an evaluation
of the background level that agrees with the estimate based on simulation studies.
The genuine dielectron background 
from WZ and ZZ production is estimated from simulation.
The misidentified electron backgrounds originate from QCD multijet and
W+jet events. These sources of background are relatively small because of the tight electron
identification and kinematic requirements, and are estimated from data
based on the probability that jets
 or random energy deposits in the calorimeters emulate electron candidates~\cite{Zprime}.

The background estimates in the dimuon and dielectron channels are tabulated in
Section~\ref{sec:unfolding} (Tables~\ref{tab:mumu-yields} and~\ref{tab:ee-yields}, respectively).

\section{Detector Resolution Effects and Unfolding}
\label{sec:unfolding}

The effects of the detector
resolution on the observed dilepton spectra are corrected through an
unfolding procedure. The original invariant mass spectrum is
related to the observed one (in the limit of no background) by
\begin{equation}
   N_{{\mathrm{obs},i}} = \sum_k \, T_{ik} \, N_{\mathrm{true},k} ,
\end{equation}
where $N_i$ is the event count in a given invariant mass bin~$i$. 
The element $T_{ik}$ of the ``response matrix'' $T$ is the probability 
that an event with an original invariant mass in the bin $k$ is reconstructed 
with an invariant mass in the bin $i$. 
The original invariant mass spectrum is obtained by inverting the response 
matrix and calculating~\cite{Cowan-unfolding,Bohm-unfolding} 
\begin{equation}
     N_{\text{u},k} \equiv N_{\mathrm{true},k} = \sum_i \, (T^{-1})_{ki} \, N_{{\mathrm{obs}},i} .
\label{eq:invResponse}
\end{equation}
This procedure is sufficient in the analysis reported in
this paper 
because the response matrix is nonsingular and nearly diagonal. 
Two extra dilepton invariant mass bins are included in the unfolding procedure, 
to account for events observed with $\MLL < 15$~\GeV or 
$\MLL > 600$~\GeV.

The response matrix is calculated using the simulated sample of DY 
events, defining the ``true mass'' as the ``generator level'' dilepton invariant mass, 
after 
FSR. Only the 
selected
events in the sample are used to calculate the response matrix. The loss of events
caused by reconstruction inefficiencies or limited acceptance is
factored out from the unfolding procedure and taken into account
by means of efficiency and acceptance factors in a 
subsequent step. Events generated with a dilepton invariant mass in the window
of the analysis but reconstructed with an invariant mass too small (below
15~\GeV) or too large (above 600~\GeV) contribute to the response
matrix. Events generated outside this window but reconstructed inside it 
are also accounted for. The sum of probabilities in the columns of the response
matrix plus the probabilities of the bins with too small and too large invariant 
masses is constrained to be 100\%.

\begin{figure}[hbtp]
  \begin{center}
    \includegraphics[width=0.49\textwidth]{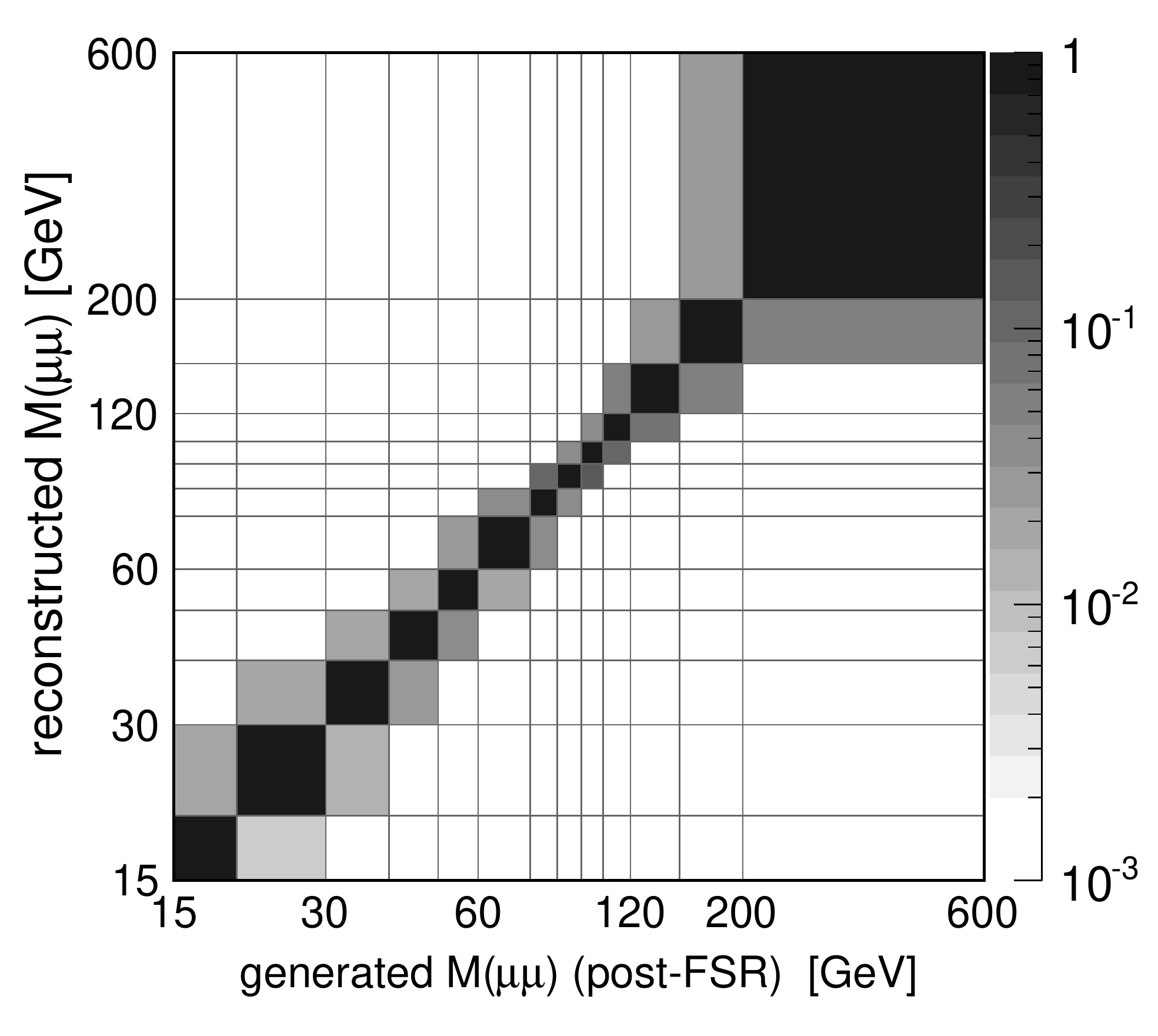}
    \includegraphics[width=0.49\textwidth]{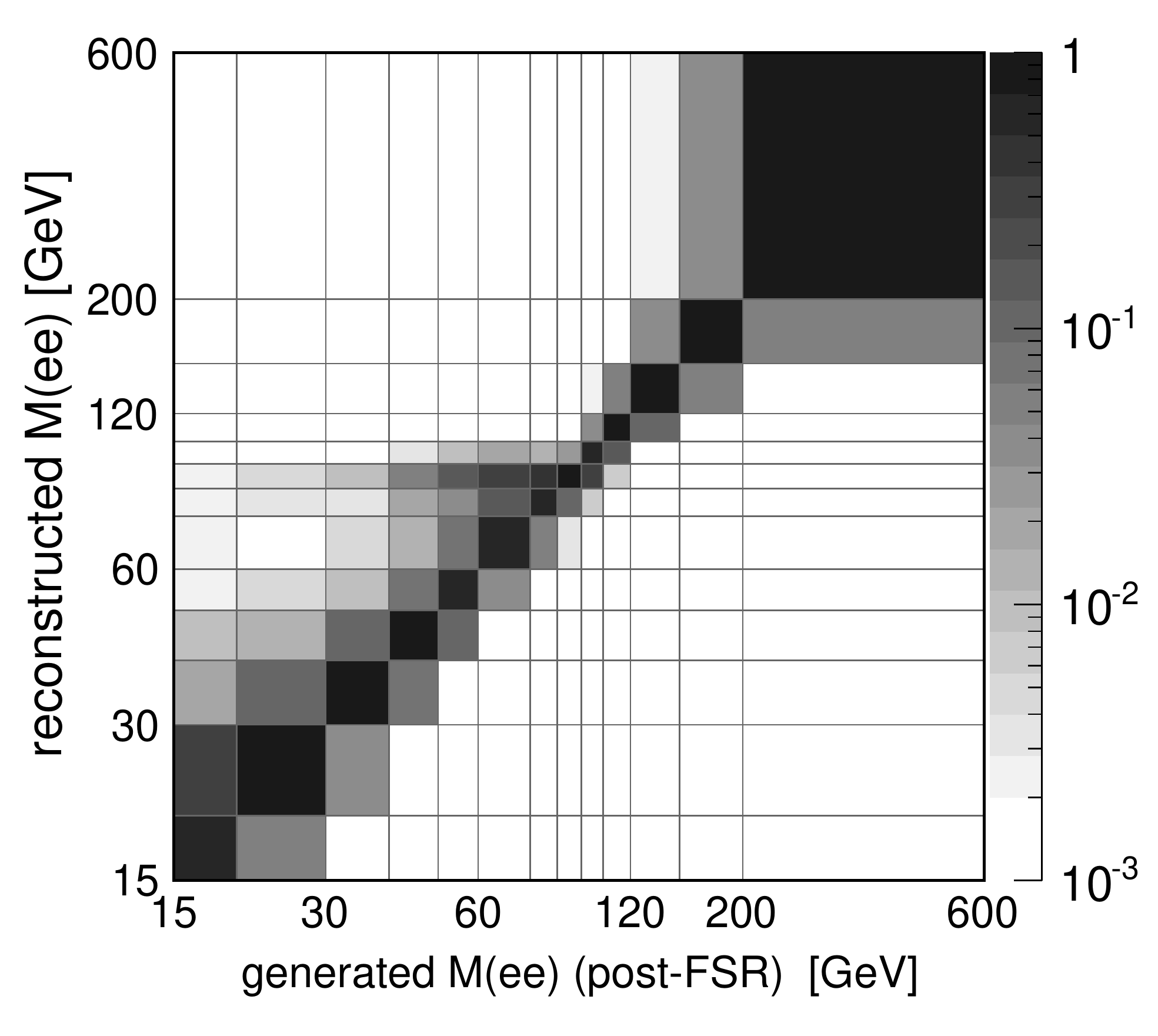}
    \caption{The response matrices for the muon (left) and electron (right) 
      channels from simulation.
       \label{fig:response-matrix}
    }
  \end{center}
\end{figure}

The response matrices are nearly diagonal. The few significant
off-diagonal elements present are found immediately next to the
diagonal elements. Almost all off-diagonal elements are less than 0.1
for the muon channel and less than 0.3 for the electron channel, as
shown in Fig.~\ref{fig:response-matrix}. The response matrices in
both lepton channels are invertible.

The larger off-diagonal elements in the response matrix for the
electron channel reflect a larger crossfeed among neighboring bins
due to the following two factors. First, the detector resolution is
worse for electrons than for muons. 
Second, the electron reconstruction algorithm attributes the four-momenta of some 
FSR photons to the electrons.  Thus, for electrons, unfolding removes not only 
the effect of detector resolution on the invariant mass but also the effect of FSR 
photons in the electron reconstruction, yielding the original mass spectrum after FSR.
The calculation of the
original mass spectrum before FSR from the spectrum resulting from the
unfolding procedure is done in a separate step through FSR corrections
and is described in the next section.

The yields before and after background subtraction and the unfolding corrections
are given in Tables~\ref{tab:mumu-yields} and~\ref{tab:ee-yields}.

\begin{table}[htbH]
\begin{center}
\caption{Observed data yields, estimated backgrounds, and
           background-corrected and unfolded signal yields for
	              DY production in the $\mu^+\mu^-$ channel. The QCD background is estimated from data 
		      whereas the ``Other'' background contributions (as indicated in Fig.~\ref{fig:mass-observed}) are based on simulation. 
		      \label{tab:mumu-yields}}
\begin{tabular}{|l|r@{$~\pm~$}l|r@{$~\pm~$}l|r@{$~\pm~$}l|r@{$~\pm~$}l|r@{$~\pm~$}l|r@{$~\pm~$}l|}
\hline
Invariant mass 
& \multicolumn{2}{c|}{$N_{\text{obs}}$}
& \multicolumn{4}{c|}{Backgrounds}
& \multicolumn{2}{c|}{$N_{\text{obs}}-N_{\text{bg}}$}
& \multicolumn{2}{c|}{$N_{\text{u}}$}\\
bin (\!\GeV)  
& \multicolumn{2}{c|}{}
& \multicolumn{2}{c|}{QCD}
& \multicolumn{2}{c|}{Other}
& \multicolumn{2}{c|}{}
& \multicolumn{2}{c|}{} \\
\hline
15--20   & 253 & 16    & 11 & 8 & 1 & 1 & 241 & 18   & 243 & 19\\
20--30   & 809 & 28    & 59 & 21& 15 & 4& 735 & 36   & 736 & 37\\
30--40   & 986 & 31    & 46 & 15& 30 & 6& 910 & 36   & 907 & 37\\
40--50   & 684 & 26    & 22 & 8 & 30 & 6& 632 & 29   & 631 & 30 \\
50--60   & 471 & 22    & 11 & 7 & 25 & 6& 435 & 24   & 436 & 26 \\
60--76   & 797& 28     & 7 & 6  & 22 & 5 & 768& 29    & 752& 31\\
76--86   & 1761& 42    &\multicolumn{2}{l|}{}& 6 & 3  & 1755& 42   & 1471& 49 \\
86--96   & 11786& 109  &\multicolumn{2}{l|}{}& 25 & 6 & 11761& 109 & 12389& 119 \\
96--106  &  909& 30    &\multicolumn{2}{l|}{}& 5 & 3  & 904& 30    &  591& 38\\
106--120 & 194& 14     &\multicolumn{2}{l|}{}& 3 & 2  & 191& 14    & 178& 17\\
120--150 & 145& 12     &\multicolumn{2}{l|}{}& 4 & 3  & 141& 12    & 142& 13\\
150--200 & 53& 7       &\multicolumn{2}{l|}{}& 4 & 3  & 49& 8      & 47& 9 \\
200--600 & 30& 6       &\multicolumn{2}{l|}{}& 3 & 2  & 27& 6      & 28& 6\\
\hline
\end{tabular}
\end{center}
\end{table}

\begin{table}[htbH]
\begin{center}
\caption{Observed data yields, estimated backgrounds, and
           background-corrected and unfolded signal yields for
	              DY production in the $\ee$ channel.
		      \label{tab:ee-yields}}
\begin{tabular}{|l|r@{$~\pm~$}l|r@{$~\pm~$}l|r@{$~\pm~$}l|r@{$~\pm~$}l|r@{$~\pm~$}l|r@{$~\pm~$}l|}
\hline
Invariant mass 
& \multicolumn{2}{c|}{$N_{\text{obs}}$}
& \multicolumn{4}{c|}{Backgrounds}
& \multicolumn{2}{c|}{$N_{\text{obs}}-N_{\text{bg}}$}
& \multicolumn{2}{c|}{$N_{\text{u}}$}\\
bin (\!\GeV)  
& \multicolumn{2}{c|}{}
& \multicolumn{2}{c|}{genuine $\ee$}
& \multicolumn{2}{c|}{misidentified $\ee$}
& \multicolumn{2}{c|}{}
& \multicolumn{2}{c|}{} \\
\hline  
15--20   &       16 &   4  &    0.0 & 0.2  &  0.4& 0.7   &   16 &   4 &   16 &   6\\
20--30   &       91 &  10  &    2.5 & 1.7  &  0.9& 1.1   &   88 &  10 &   94 &  12\\
30--40   &      179 &  13  &   14.3 & 4.6  &  1.5& 1.4   &  163 &  14 &  164 &  17 \\
40--50   &      243 &  16  &   31.4 & 6.9  &  3.7& 2.7   &  208 &  18 &  219 &  22 \\
50--60   &      211 &  15  &   19.9 & 5.2  &  3.9& 2.8   &  187 &  16 &  234 &  25\\
60--76   &      455 &  21  &   22.4 & 5.3  &  4.9& 3.3   &  428 &  22 &  620 &  45\\
76--86   &     1599 &  40  &    8.5 & 2.8  &  2.5& 2.1   & 1588 &  40 & 1277 &  89\\
86--96   &     6998 &  84  &   12.5 & 1.8  &  4.4& 3.1   & 6981 &  84 & 7182 & 117\\
96--106  &      587 &  24  &    3.5 & 1.8  &  2.1& 1.8   &  581 &  24 &  441 &  36\\
106--120 &      132 &  11  &    3.2 & 1.9  &  1.5& 1.4   &  127 &  12 &  127 &  15\\
120--150 &       67 &   8  &    7.8 & 3.1  &  2.0& 1.7   &   57 &   9 &   53 &  10 \\
150--200 &       34 &   6  &    5.5 & 2.5  &  1.6& 1.4   &   27 &   7 &   25 &   7\\
200--600 &       26 &   5  &    3.0 & 1.9  &  1.4& 1.4   &   22 &   6 &   21 &   5 \\
\hline
\end{tabular}
\end{center}
\end{table}

\section{Acceptance and Efficiency}
\label{sec:AccepEff}

The reconstructed dilepton invariant mass distributions cannot be directly compared to the spectra
provided by the theoretical models, not only because of the limited acceptance coverage of the 
detector but also because the observed spectra are affected by FSR, a process usually not 
included in the calculations. We define ``pre-FSR'' and ``post-FSR'' as labels to be attached to any quantity
referred to before and after the FSR effects occur.  
The measurement of $d\sigma/d\MLL$ therefore requires a two-step correction procedure.  First, 
the measured, post-FSR spectra are corrected for acceptance, when applicable, and detector efficiencies.  
Then the (acceptance and) efficiency corrected spectra are themselves altered by a bin-by-bin FSR 
correction factor which relates the yields before and after the FSR takes place. 
These spectra can be compared to the calculations.

The geometrical and kinematic acceptance $A$ is defined, using the simulated leptons after the
FSR simulation, as $A \equiv \NACC/\NGEN$, where $\NGEN$ is the number of generated events and
$\NACC$ is the corresponding number of events passing the standard $\PT$ and $\eta$ 
lepton requirements, in each dilepton invariant mass bin.

The efficiency $\epsilon$ is the fraction of events within the acceptance that pass the full 
selection, so that
\begin{equation}\label{eqn:AccEff}
    A \cdot \epsilon \equiv \frac{\NACC}{\NGEN} \cdot
    \frac{\NEFF}{\NACC} = \frac{\NEFF}{\NGEN} ,
\end{equation}
where $\NEFF$ is the number of events surviving the reconstruction, selection, and identification requirements.
The values of the product of acceptance and efficiency are obtained from simulation. 
A separate 
correction factor is determined from data and applied to the product, following the procedure used in the inclusive W 
and Z cross section measurements in CMS~\cite{ZCrossSection}. This factor, the efficiency correction,
describes the difference 
between data and simulation
in the efficiency to observe single leptons or dileptons.

The {\sc POWHEG} simulation combines the next-to-leading-order (NLO) calculations with a parton showering which 
is insufficient to model fully the low invariant mass region of the dilepton spectra.
The two high-\pt leptons required in the analysis must form a small angle at low mass and therefore the dilepton 
system gets significantly boosted, something to be compensated by hard gluon radiation in the transverse plane. This means 
that these low-mass events are of the type ``$\gamma^*$ + hard jet'' at first order, and therefore the next order of 
correction (NNLO) becomes essential for a reliable estimate of acceptance corrections.
To account for this, a
correction is applied, determined from the ratio between the differential cross sections calculated
at NNLO with {\sc FEWZ}~\cite{FEWZ} and at NLO with {\sc POWHEG}, both at pre-FSR level.  These correction weights, obtained
in bins of dilepton rapidity, \pt, and invariant mass, are applied on an event-by-event basis.
The distributions obtained are used for all the simulation based estimations (acceptance, efficiency, FSR corrections) for DY, and this sample is 
referred to as ``{\sc POWHEG} matched to {\sc FEWZ} (NNLO) distributions''. 
This procedure changes the acceptance in the lowest invariant mass bin significantly (by about 50\%), 
but has a small effect, not exceeding 3\%, on the rest of the bins.

\begin{figure}[h]
{\centering
\includegraphics[width=0.365\textwidth, angle = 90]{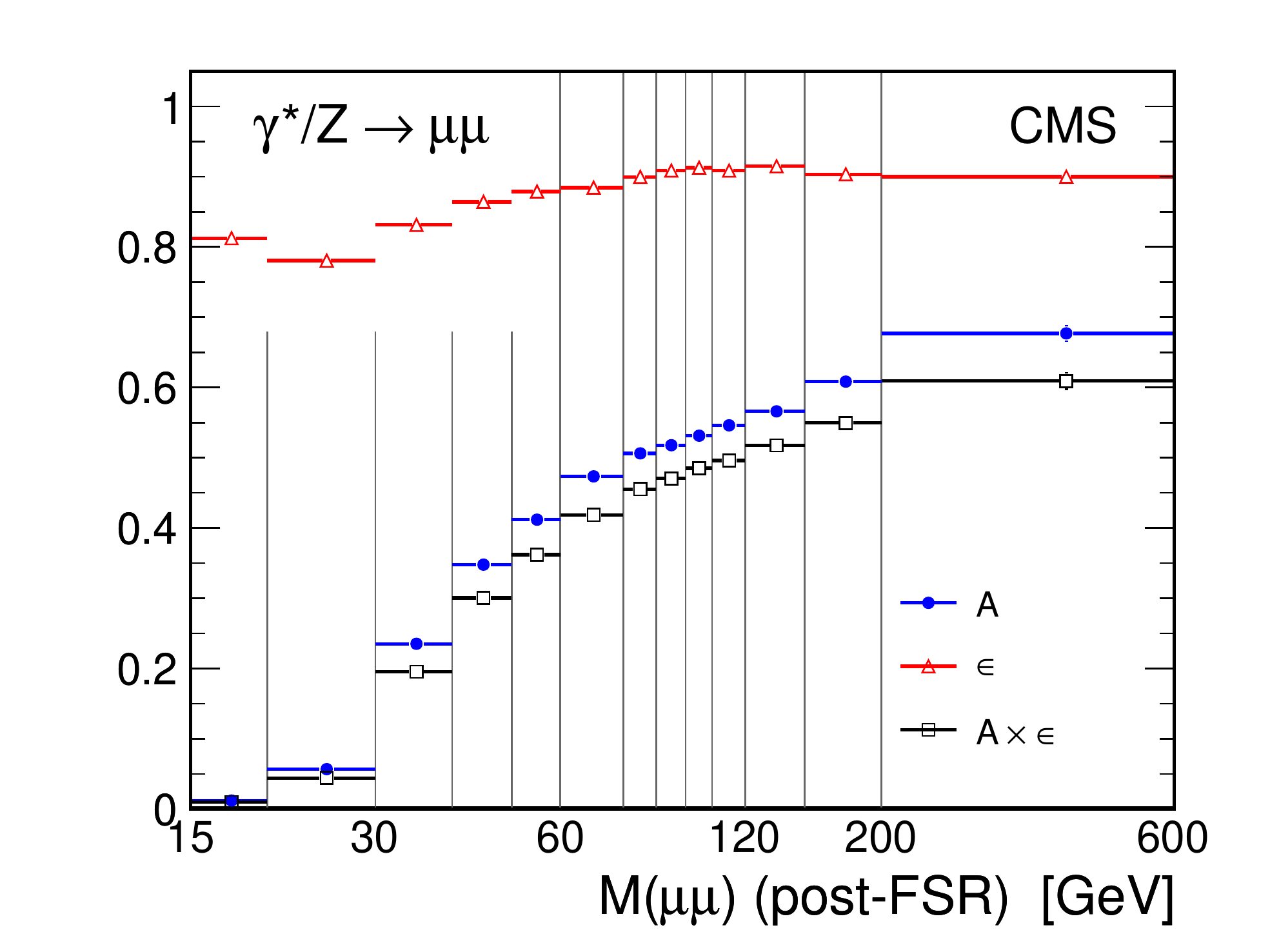}
\includegraphics[width=0.365\textwidth, angle = 90]{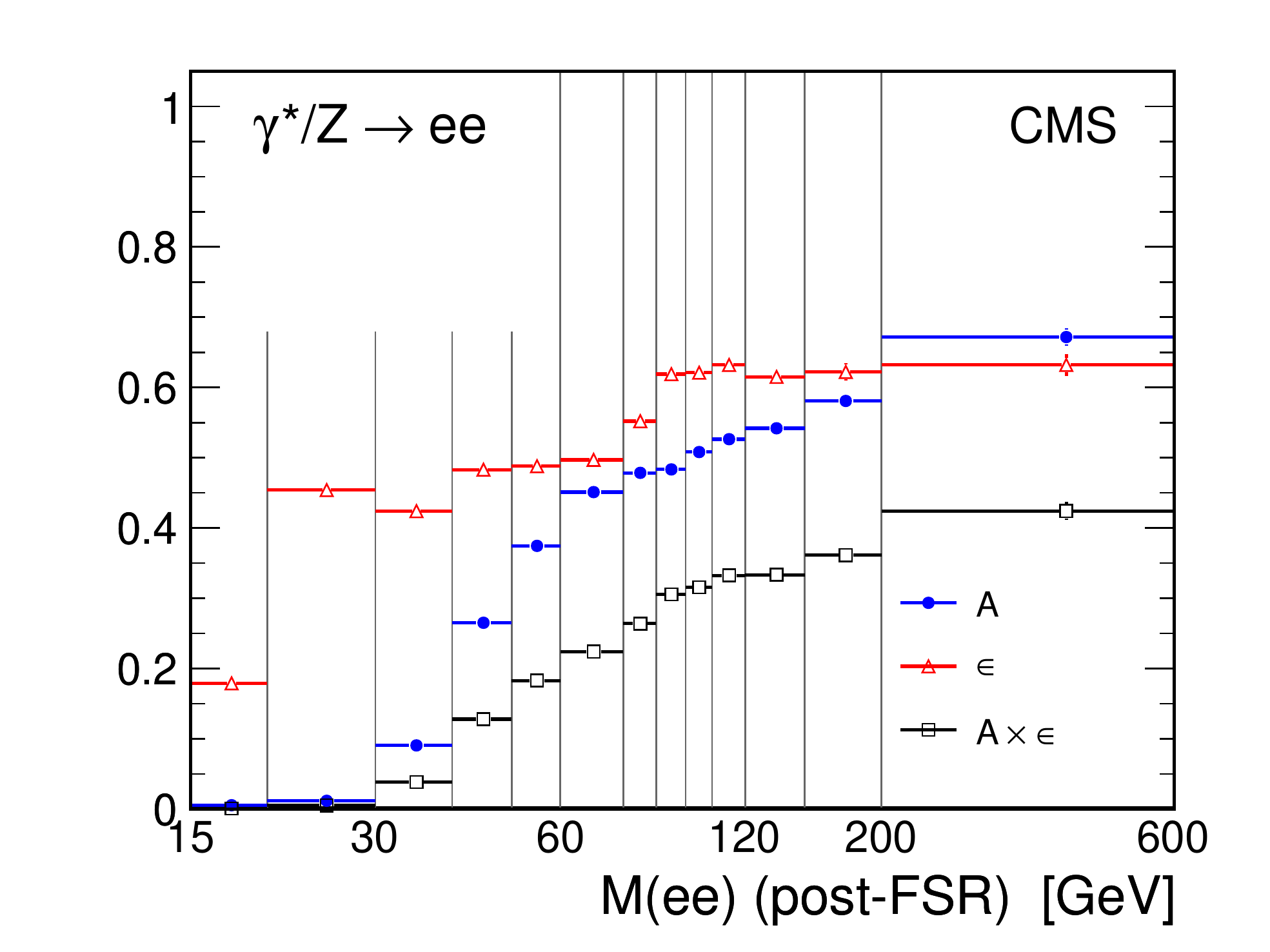}
\caption{\label{Acc}
DY acceptance (blue, filled circles), efficiency (red, open triangles), and their product (black, open squares) per invariant mass bin, 
for the $\mu^+\mu^-$ (left) and $\ee$ (right) channels.}
}
\end{figure}

Figure~\ref{Acc} shows the variables $A$, $\epsilon$, and
$A\cdot\epsilon$ as functions of $\MLL$ for dimuons (left) and
dielectrons (right), the values being listed in
Tables~\ref{tab_accEff} and~\ref{tab_accEff_electrons}, respectively.

The FSR correction factors listed in
Tables~\ref{tab_accEff} and~\ref{tab_accEff_electrons} for
a given invariant mass range are obtained 
from simulation by dividing the post-FSR cross sections by the corresponding pre-FSR 
quantities. They are applied on (corrected) data as an additional step as described earlier
in the section.
The factors obtained within the detector acceptance and in the full phase space 
(as shown in the tables) are applied to the corresponding measurements.
Systematic uncertainties related to the FSR simulation are discussed in 
Section~\ref{sec:systematics}.

\begin{table}[h]
\begin{center}
\caption{DY acceptance and acceptance times efficiency per invariant mass bin
for the $\mu^+\mu^-$ channel.
In addition, the FSR correction factors are given. All uncertainties are statistical.\label{tab_accEff} }
\begin{tabular}{|l|r@{$~\pm~$}l|r@{$~\pm~$}l|r@{$~\pm~$}l|r@{$~\pm~$}l|}
\hline
Invariant mass 
& \multicolumn{2}{c|}{Acceptance (\%)}
& \multicolumn{2}{c|}{Acc $\times$ Eff (\%)}
& \multicolumn{2}{c|}{FSR correction (\%)}
& \multicolumn{2}{c|}{FSR correction in}\\
bin (\!\GeV) 
& \multicolumn{2}{c|}{}
& \multicolumn{2}{c|}{}
& \multicolumn{2}{c|}{}
& \multicolumn{2}{c|}{the acceptance (\%)} \\
\hline

15--20 & 1.23 & 0.01 & 1.00 & 0.01     &  97.28 & 0.02 & 96.30 & 0.02 \\
\hline
20--30 & 5.69 & 0.03 & 4.44 & 0.03     &  97.28 & 0.02 & 97.99 & 0.02 \\
\hline
30--40 & 23.5 & 0.1 & 19.6 & 0.1   &  98.43 & 0.03 & 98.77 & 0.03 \\
\hline
40--50 & 34.8 & 0.2 & 30.1 & 0.2   &  104.0 & 0.1 & 105.9 & 0.1 \\
\hline
50--60 & 41.2 & 0.2 & 36.2 & 0.2   &  120.2 & 0.3 & 125.1 & 0.3 \\
\hline
60--76 & 47.4 & 0.2 & 41.9 & 0.2   & 166.4 & 0.5 & 175.1 & 0.6 \\
\hline
76--86 & 50.6 & 0.1 & 45.5 & 0.1   & 167.1 & 0.4 & 169.8 & 0.4 \\
\hline
86--96 & 51.8 & 0.1 & 47.1 & 0.1   & 91.63 & 0.03 & 91.62 & 0.03 \\
\hline
96--106 & 53.1 & 0.2 & 48.5 & 0.2  &  88.0 & 0.1 & 88.1 & 0.1 \\
\hline
106--120 & 54.6 & 0.4 & 49.6 & 0.4 &  91.3 & 0.2 & 91.2 & 0.2 \\
\hline
120--150 & 56.6 & 0.6 & 51.8 & 0.6 &  93.2 & 0.3 & 93.1 & 0.3 \\
\hline
150--200 & 60.8 & 0.9 & 55.0 & 0.9 &  94.3 & 0.4 & 95.0 & 0.4 \\
\hline
200--600 & 67.7 & 1.2 & 60.9 & 1.3 &  92.8 & 0.7 & 93.1 & 0.6 \\
\hline
\end{tabular}
\end{center}
\end{table}

\begin{table}[h]
\begin{center}
\caption{DY acceptance and acceptance times efficiency per invariant mass bin
for the $\ee$ channel. 
In addition, the FSR correction factors are given. All uncertainties are statistical.\label{tab_accEff_electrons} }
\begin{tabular}{|l|r@{$~\pm~$}l|r@{$~\pm~$}l|r@{$~\pm~$}l|r@{$~\pm~$}l|}
\hline
Invariant mass 
& \multicolumn{2}{c|}{Acceptance (\%)}
& \multicolumn{2}{c|}{Acc $\times$ Eff (\%)}
& \multicolumn{2}{c|}{FSR correction (\%)}
& \multicolumn{2}{c|}{FSR correction in}\\
bin (\!\GeV) 
& \multicolumn{2}{c|}{}
& \multicolumn{2}{c|}{}
& \multicolumn{2}{c|}{}
& \multicolumn{2}{c|}{the acceptance (\%)} \\
\hline
15--20   &  0.56 & 0.01 &   0.10 & 0.01 &   93.8 & 0.1 &   98.7 & 1.9  \\ \hline
20--30   &  1.19 & 0.01 &   0.54 & 0.01 &   93.9 & 0.2 &  102.9 & 1.8  \\ \hline
30--40   &  9.1 & 0.1 &   3.6 & 0.1 &   96.8 & 0.3 &  109.5 & 1.3  \\ \hline
40--50   &  26.5 & 0.2 &  12.8 & 0.2 &  107.7 & 0.6 &  117.8 & 1.2  \\ \hline
50--60   &  37.5 & 0.2 &  18.3 & 0.2 &  139.3 & 1.0 &  156.2 & 1.8  \\ \hline
60--76   &  45.1 & 0.2 &  22.4 & 0.2 &  230.7 & 1.4 &  256.3 & 2.4  \\ \hline
76--86   &  47.9 & 0.1 &  26.4 & 0.1 &  224.1 & 1.0 &  235.0 & 1.5  \\ \hline
86--96   &  49.3 & 0.1 &  30.6 & 0.1 &   83.9 & 0.1 &   85.6 & 0.2  \\ \hline
96--106  &  50.8 & 0.2 &  31.6 & 0.2 &   78.5 & 0.5 &   80.1 & 0.7  \\ \hline
106--120 &  52.6 & 0.4 &  33.3 & 0.4 &   83.9 & 1.0 &   85.2 & 1.4  \\ \hline
120--150 &  54.2 & 0.6 &  33.4 & 0.6 &   87.9 & 1.4 &   88.5 & 1.9  \\ \hline
150--200 &  58.1 & 0.9 &  36.1 & 0.9 &   89.1 & 2.2 &   90.3 & 3.0  \\ \hline
200--600 &  67.2 & 1.3 &  42.4 & 1.3 &   87.5 & 3.2 &   88.9 & 4.0  \\ \hline
\end{tabular}
\end{center}
\end{table}

The total dimuon event selection efficiency is factorized as

\begin{equation}
\varepsilon(\text{event}) = 
       \varepsilon(\mu_1)
       \cdot \varepsilon(\mu_2)
       \cdot \varepsilon[\mu\mu|(\mu_1) \& (\mu_2)]
       \cdot \varepsilon(\text{event},\text{trig}|\mu\mu) ,
\label{eq:muon-event-eff}
\end{equation}

where $\varepsilon(\mu)$ is the single muon selection efficiency;
$\varepsilon[\mu\mu|(\mu_1) \& (\mu_2)]$ is the dimuon 
selection efficiency, which includes 
the requirement that 
the two muon tracks be consistent with originating from a common vertex and that they satisfy the angular criteria;
and
$\varepsilon(\text{event},\text{trig}|\mu\mu)$ is the efficiency of triggering an event 
including the efficiency that an identified muon is matched to a trigger object.
The single muon efficiency is factorized as

\begin{equation}
\varepsilon(\mu)= \varepsilon(\text{track}|\text{accepted})
                             \cdot\varepsilon(\text{reco}+\text{id}|\text{track})
                             \cdot\varepsilon(\text{iso}|\text{reco}+\text{id}) ,
\label{eq:single-muon-eff}
\end{equation}

where 
$\varepsilon(\text{track}|\text{accepted})$ is the offline track reconstruction efficiency in the tracker detector;
$\varepsilon(\text{reco}+\text{id}|\text{track})$ is the muon reconstruction and identification efficiency; and
$\varepsilon(\text{iso}|\text{reco}+\text{id})$ is the muon isolation efficiency.
The trigger efficiency $\varepsilon(\text{event},\text{trig}|\mu\mu)$ is given by

\begin{equation}
\varepsilon(\text{event},\text{trig}|\mu\mu) 
     = \varepsilon(\mu_1,\text{trig}|\mu_1) 
     + \varepsilon(\mu_2,\text{trig}|\mu_2)
     - \varepsilon(\mu_1,\text{trig}|\mu_1)
         \cdot\varepsilon(\mu_2,\text{trig}|\mu_2) ,
\end{equation}

where $\varepsilon(\mu,\text{trig}|\mu)$ 
is the efficiency of an offline selected muon to fire the trigger.

The track reconstruction efficiency is very high ($99.5\%$). The angular 
criterion is nearly $100\%$ efficient for signal DY events, and the vertex 
probability requirement is more than $98\%$ efficient and has a negligible 
($< 0.3\%$) dependence on~$\MLL$.

The muon reconstruction and identification efficiency is estimated using 
clean samples of muon pairs in the Z peak (tag and probe, T\&P, method~\cite{ZCrossSection}). 
The properties of one muon are probed, after imposing tight requirements 
on the other one.
To determine the isolation efficiency, the Lepton Kinematic Template Cones (LKTC) method~\cite{tag-and-probe}  is applied.
The essence of the LKTC method is to choose predefined directions in events 
with 
an underlying event environment similar to that 
of the signal sample. The isolation variable is  defined as if these directions 
represent signal leptons, and the chosen isolation-based criteria are subsequently studied.

To describe the observed efficiency variations between data and simulation, efficiency correction factors are obtained in bins of $\PT$ and $\eta$
as the ratio of the efficiencies measured with
data and with the simulated events:
\begin{equation}
\rhoeff(\PT,\eta) = \frac{\effdata(\PT,\eta)}{\effmc(\PT,\eta)} .
\label{eqn:rho-definition}
\end{equation}

The corrections to the efficiencies in simulation are implemented by reweighting simulated events, with 
weights computed as 
$W =\rho_1^{\text{reco}}  \rho_2^{\text{reco}}    \rho_1^{\text{iso}}   \rho_2^{\text{iso}} \rho^{\text{trig}}$ where 
$\rho^{\text{trig}} = (\epsilon_{\text{data},1}^{\text{trig}} +  \epsilon_{\text{data},2}^{\text{trig}} - \epsilon_{\text{data},1}^{\text{trig}}  \epsilon_{\text{data},2}^{\text{trig}})/(\epsilon_{\text{MC},1}^{\text{trig}} +  \epsilon_{\text{MC},2}^{\text{trig}} - \epsilon_{\text{MC},1}^{\text{trig}}  \epsilon_{\text{MC},2}^{\text{trig}})$.
If $\PT < 16$~\GeV or $|\eta| > 2.1$ for a given muon~ $i=1,2$, its 
trigger efficiency
is set to zero.

The systematic uncertainty related to the efficiency correction is 
evaluated by generating one hundred variations of the $(\PT,\eta)$ 
correction maps,
where the weight in each $(\PT,\eta)$ bin is
obtained by adding to the original value a Gaussian-distributed shift of
mean zero and width equal to the statistical uncertainty of the
original correction factor (Eq.~(\ref{eqn:rho-definition})). 
Signal corrected yields are evaluated using event weights
obtained from each of the alternative correction maps and the RMS
spread of the resulting values is taken as the systematic
uncertainty. 
The systematic error computed with this procedure includes an irreducible statistical component, yielding a conservative uncertainty
which also covers 
generous variations in the efficiency-correction shape.
The resulting uncertainties are shown in
Table~\ref{tab_effCorr}.

The total event efficiency in the dielectron channel analysis is defined
as the product of the two single electron efficiencies, which incorporate three
factors: 1)~the efficiency $\effreco$ to reconstruct an electron candidate from an energy
deposit in the ECAL; 2)~the efficiency $\effid$ for that candidate to pass the selection
criteria, including identification, isolation, and conversion rejection; 3)~the efficiency
$\efftrig$ for the leading electron to pass the trigger
requirements. Each of these efficiencies is obtained from simulation and corrected by
$\rhoeff(\PT,\eta)$, as for the muon channel (Eq.~(\ref{eqn:rho-definition})).  The 
T\&P method is used for all efficiency components.  The event efficiency correction
and its uncertainty are derived as for the muon channel by reweighting simulated events.
The correction factors are listed in Table~\ref{tab_effCorr}.

\begin{table}[h]
\begin{center}
\caption{Combined efficiency corrections for the muon and electron channels per mass bin.
They account for the data vs.\ simulation differences in reconstruction, identification, isolation and trigger efficiencies.
\label{tab_effCorr} }
\begin{tabular}{| l | l | l |}
\hline
Invariant mass &  \multicolumn{2}{c|}{ Combined efficiency correction} \\
bin (\!\GeV)    & Muon channel & Electron channel\\
\hline
15--20 & $0.917\pm 0.010$ & $1.098\pm 0.087$ \\
\hline
20--30 & $0.915\pm 0.010$ & $1.089\pm 0.091$\\
\hline
30--40 & $0.918\pm 0.011$ & $1.107\pm 0.103$\\
\hline
40--50 & $0.931\pm 0.011$ & $1.076\pm 0.081$\\
\hline
50--60 & $0.943\pm 0.008$ & $1.034\pm 0.053$\\
\hline
60--76 &  $0.952\pm 0.006$ & $1.008\pm 0.033$\\
\hline
76--86 &  $0.958\pm 0.004$ & $0.995\pm 0.024$\\
\hline
86--96 &  $0.960\pm 0.003$ & $0.979\pm 0.019$\\
\hline
96--106 &  $0.961\pm 0.003$ & $0.973\pm 0.018$\\
\hline
106--120 & $0.961\pm 0.003$ & $0.960\pm 0.018$\\
\hline
120--150&  $0.956\pm 0.010$ & $0.953\pm 0.019$\\
\hline
150--200&  $0.957\pm 0.021$ & $0.945\pm 0.020$\\
\hline
200--600&  $0.957\pm 0.021$ & $0.940\pm 0.020$\\
\hline
\end{tabular}
\end{center}
\end{table}

\section{Systematic Uncertainties}
\label{sec:systematics}

Systematic uncertainties have been evaluated for each step in the determination of the dilepton invariant mass
spectrum. The acceptance-related uncertainties are a special case as they 
only apply to the acceptance corrected results, i.e., results in the full phase space, 
and are approximately the same for the dimuon and dielectron channels 
(the FSR uncertainties are treated separately).
The acceptance uncertainty resulting from the knowledge of the PDFs is estimated using {\sc PYTHIA} 
with the CTEQ6.1 PDF set by a reweighting technique~\cite{Bourilkov:2006cj}, with a negligible statistical uncertainty given the very large 
simulated sample.
Since we are making a shape measurement, normalizing the DY cross
section to the dilepton cross section in the $\text{Z}$ region, the analysis only 
depends on the uncertainty of the \emph{ratio} of acceptances,
$A_i/\ANORM$, where $A_i$ is the acceptance for 
the invariant mass bin $i$ and $\ANORM$ is the acceptance for the
invariant mass region of the $\text {Z}$. 

The uncertainty of the acceptance is estimated, for each dilepton invariant mass bin,
using {\sc FEWZ}, at NLO and NNLO accuracy
in perturbative QCD.   Variations of the factorization and renormalization
scales lead to a systematic uncertainty smaller than 1\% (at NNLO) 
for most of the invariant mass range used in the analysis presented here.

Special care is needed to calculate the acceptance of low invariant mass dileptons, 
where differences between  NLO and NNLO values can be significant, 
given the relatively high thresholds imposed on the transverse momentum of
the leptons. 
Since the {\sc POWHEG} MC (NLO) simulation, modified to match the {\sc FEWZ} (NNLO) calculations, is used 
to calculate the acceptance
corrections used in the analysis, 
an additional (model-dependent) systematic uncertainty on the acceptance
calculation is determined from the 
observed differences in acceptances based on
{\sc FEWZ} spectra and {\sc POWHEG} distributions matched to {\sc FEWZ}.
These differences are caused by variations in the kinematic
distributions within the bins 
where bin sizes are chosen to take into account the limited reliability of perturbative 
QCD calculations in parts of the phase space.
This systematic uncertainty reaches up to 10\% in the dilepton
invariant mass range considered in the analysis and is included in the comparison 
between the measurements and the theoretical expectations.

The dominant systematic uncertainty on the cross section measurement in the 
dimuon channel is the uncertainty on the background estimation, which is,
however, relatively small given the low background levels. This uncertainty is
evaluated from data using two independent background subtraction methods, as described in Section~\ref{sec:backgrounds}.
The next most important uncertainties are related to 
the muon efficiency and to the muon momentum scale and resolution. 
The former is determined using the large sample of Z events decaying to
dimuons. 
Uncertainties in the latter are mostly caused by 
residual misalignment between the muon chambers and the silicon tracker,
potentially not reproduced in the simulation. The Z line shape is used to 
constrain the level of such possible limitations in the simulation.
The momentum resolution and the momentum scale uncertainties are included
in the unfolding procedure and, hence, the resulting shape is affected
by these systematic effects.
The level of the momentum scale uncertainty is evaluated by introducing a  bias in the MC reconstruction and unfolding the resulting dimuon mass distribution with the unfolding matrix determined from the nominal (unbiased) MC sample. The bias is on the reconstructed invariant mass and is based on the maximal difference between MC and data Z peak positions as obtained with variations in the \pt and $\eta$ requirements.

Studies of photons reconstructed near a muon in a DY event indicate that 
the FSR simulation is remarkably accurate.  A corresponding systematic uncertainty
is evaluated by examining how 
the results change when the fraction of FSR events as well as the energy and 
angular distributions of the radiated photon are 
modified within proper statistical variations.

Other systematic effects that could affect the dimuon yield have been considered, 
such as the impact of 
additional soft pp collisions that occur in the same bunch crossing as
the studied interaction 
and the effects of the dimuon vertex 
probability requirement and of residual data-simulation discrepancies.  A combined uncertainty is
reported for these ``other'' sources in Table~\ref{tab_syst}, where all systematic 
uncertainties in the dimuon channel are listed.

\begin{table}[h]
\begin{center}
\caption{Summary of systematic uncertainties in the muon channel (in percent). The ``Total'' is 
a quadratic sum of all sources without ``Acceptance''. With the exception 
of ``Acceptance'', the numbers correspond to the individual measurements per bin and not the 
ratio to the Z region.
\label{tab_syst} }
\begin{tabular}{| l | l | l | l | l | l | l || l |}
\hline
Invariant mass&Efficiency& Background& Unfolding & FSR &Other & Total&Acceptance\\
bin (\!\GeV) & correction && & &  & & \\
\hline
15--20 & $1.1 $    & $3.6$  & $0.4$ & $1.5 $  &$1.0$ &$4.2$ & $+2.2$/$-3.0$\\
20--30 & $1.1 $  & $3.1$  & $0.2$  & $1.1 $  &$1.0$ &$3.6$  & $+1.9$/$-3.2$\\
30--40 & $1.2 $  & $1.9$  & $0.1$  & $0.7 $  &$1.0$ &$2.6$  & $+1.7$/$-3.0$\\
40--50 & $1.2$   & $1.7$  & $0.2$  & $0.7 $  &$1.0$ &$2.4$& $+1.7$/$-2.9$\\
50--60 & $0.8$   & $2.1$  & $0.2$  & $0.5 $  &$0.5$ &$2.4$& $+1.7$/$-2.8$\\
60--76 & $0.6$   & $1.0$  & $0.2$  & $1.4 $  &$0.5$ &$1.9$& $+1.6$/$-2.6$\\
76--86 & $0.4$   & $0.2$  & $1.7$  & $2.0 $  &$0.5$ &$2.7$& $+1.5$/$-2.5$\\
86--96 & $0.3$   & $0.05$ & $0.2$  & $0.5 $  &$0.5$ &$0.8$& $+1.5$/$-2.4$\\
96--106 &  $0.3$ & $0.4$  & $3.8$  & $0.5 $  &$0.5$ &$3.9$& $+1.5$/$-2.4$\\
106--120 & $0.3$ & $1.4$    & $0.7$  & $0.5 $  &$3.0$ &$3.4$& $+1.5$/$-2.3$\\
120--150& $1.1$  & $2$    & $0.4$  & $0.5 $  &$1.0$ &$2.6$& $+1.5$/$-2.1$\\
150--200& $2.1$  & $6$    & $0.9$  & $0.5 $  &$1.0$ &$6.5$  & $+1.4$/$-1.8$\\
200--600& $2.1$  & $10$   & $0.1$  & $0.5 $  &$1.0$ &$10.3$ & $+1.2$/$-1.4$\\
\hline
\end{tabular}
\end{center}
\end{table}

\begin{table}[h]
\begin{center}
\caption{Summary of systematic uncertainties in the electron channel (in percent). The ``Total'' is 
a quadratic sum of all sources without ``Acceptance''. With the exception 
of ``Acceptance'', the numbers correspond to the individual measurements per bin and not the
ratio to the Z region.
\label{tab_syst_ee} }
\begin{tabular}{| l | l | l | l | l | l || l |}
\hline
Invariant mass& Energy & Efficiency& Background& Unfolding & Total &Acceptance\\
bin (\!\GeV) & scale &correction & & &  &  \\ 
\hline
15--20  & $ 23.4$ &  $  9.2$ &  $  6.2$ &  $  8.7$ &  $ 27.3$  & $+2.1$/$-2.9$\\ 
20--30  & $  3.6$ &  $  8.5$ &  $  2.8$ &  $  2.1$ &  $  9.9$  & $+1.7$/$-2.8$\\ 
30--40  & $  2.7$ &  $  9.4$ &  $  4.0$ &  $  1.5$ &  $ 10.6$  & $+1.5$/$-2.7$\\ 
40--50  & $  3.3$ &  $  7.5$ &  $  5.2$ &  $  1.4$ &  $  9.9$  & $+1.5$/$-2.5$\\ 
50--60  & $  3.3$ &  $  5.2$ &  $  4.6$ &  $  1.9$ &  $  7.9$  & $+1.5$/$-2.4$\\ 
60--76  & $ 10.3$ &  $  3.3$ &  $  2.2$ &  $  2.0$ &  $ 11.2$  & $+1.4$/$-2.3$\\ 
76--86  & $ 39.5$ &  $  2.5$ &  $  0.8$ &  $  3.1$ &  $ 39.7$  & $+1.3$/$-2.2$\\ 
86--96  & $  3.9$ &  $  1.9$ &  $  0.2$ &  $  0.6$ &  $  4.4$  & $+1.2$/$-2.1$\\ 
96--106 & $ 45.6$ &  $  2.0$ &  $  0.9$ &  $  3.6$ &  $ 45.8$  & $+1.3$/$-2.0$\\
106--120& $ 13.2$ &  $  2.1$ &  $  2.6$ &  $  2.4$ &  $ 13.9$  & $+1.3$/$-1.9$\\
120--150& $  6.0$ &  $  2.4$ &  $  8.2$ &  $  2.6$ &  $ 10.8$  & $+1.3$/$-1.8$\\ 
150--200& $  5.7$ &  $  2.8$ &  $ 12.9$ &  $  2.4$ &  $ 14.5$  & $+1.2$/$-1.5$\\
200--600& $  4.6$ &  $  3.2$ &  $ 11.8$ &  $  1.6$ &  $ 13.1$  & $+1.0$/$-1.1$\\
\hline
\end{tabular}
\end{center}
\end{table}

In the electron channel, the leading systematic uncertainty is
associated with the energy scale corrections of individual electrons.
The corrections affect both the placement of a given candidate in a particular
invariant mass bin and the likelihood of surviving the kinematic
selection. The energy scale correction itself is calibrated to 2\%
precision for the dataset used. The associated error on signal event
yields is calculated by varying the energy scale correction value
within this amount and remeasuring the yields. This uncertainty takes
its largest values for the bins just below and above the central
Z peak bin because of bin migration. The energy scale uncertainty for the
electron channel is on the order of 20 times larger than the momentum scale
uncertainty for muons, for which the associated systematic uncertainties on the
cross section are rather small.

The second leading uncertainty for electrons is caused by the uncertainty on the
efficiency scale factors. The precision of the scale factor calibration is
limited by the size of the data sample available for the T\&P procedure.
The systematic uncertainty on the scale factors as well as the
resulting error on the normalized cross section are found with the
same procedure as for the muon channel.

The dielectron background uncertainties are evaluated by comparing the
background yields calculated as described in
Section~\ref{sec:backgrounds} with predictions from simulation. These
uncertainties are only dominant at the highest invariant masses
considered. The uncertainty associated with the unfolding procedure in
the electron channel comes primarily from the uncertainty on the
unfolding matrix elements due to imperfect simulation of detector
resolution. This simulation uncertainty for electrons is significantly 
larger than for muons, leading to a larger systematic uncertainty on 
the normalized cross section. The uncertainties due to FSR effects 
are estimated with a method similar to that for the muon channel discussed
above with similar values. Because of significantly higher
systematic uncertainty for all mass bins for the electron channel than
for the muon channel, the FSR related contribution to the electron channel
systematic uncertainty is neglected. 

The systematic uncertainties for the electron channel are summarized in 
Table~\ref{tab_syst_ee}. At present the dominant systematic uncertainties 
are driven by the limited size of calibration samples available for energy 
scale and efficiency scale factor calculations, and therefore the 
uncertainties could be reduced significantly with larger data samples.

\section{Results}
\label{sec:results}

The DY cross section per invariant mass bin $i$, $\sigma_i$, is calculated according to Eq. (\ref{eqn:fullCrossSection_intro}).

In order to provide a measurement independent of the luminosity uncertainty and to reduce
many systematic uncertainties, the $\sigma_i$ is normalized to
the cross section in the Z region, $\sigma_{\mathrm{\ell\ell}}$, defined as the DY 
cross section in the invariant mass region $60 < \MLL < 120~\GeV$.  
The result of the analysis is presented as the ratio
\begin{equation}
\label{eqn:fullCrossSectionRatio}
  R^i_{\text{post-FSR}} = \frac{N_{\text{u},i}}{A_i\,\varepsilon_i\,\rho_i} \big/
   \frac{\NUNORM}{\ANORM\,\effNORM\,\rhoNORM},
\end{equation}
where $N_{\text{u},i}$ is the number of events after the unfolding procedure, and the
acceptances $A_i$, the efficiencies $\epsilon_i$, and the corrections estimated from data,
$\rho_i$, were defined earlier; $\NUNORM$, $\ANORM$, $\effNORM$, and $\rhoNORM$ 
refer to the Z region.
For both lepton channels, the cross sections 
in the Z region measured in this analysis are in excellent agreement with the 
previous CMS measurement~\cite{ZCrossSection}.

In order to allow a more direct and precise comparison with theory predictions, the 
shape measured before the acceptance correction is also reported, thus eliminating PDF and theory 
uncertainties from the experimental results:
\begin{equation}
\label{eqn:fullCrossSectionRatio_DET}
  R_{\text{det, post-FSR}}^i = \frac{N_{\text{u},i}}{\varepsilon_i\,\rho_i} \big/
   \frac{\NUNORM}{\effNORM\,\rhoNORM} .
\end{equation}

The post-FSR shapes, $\RPOSTFSR$ and $\RPOSTFSRDET$, are modified by the FSR correction factors from Tables~\ref{tab_accEff} and~\ref{tab_accEff_electrons} 
to obtain the pre-FSR shapes, $R$ and $\RDET$, respectively.
The shapes integrated in the normalization region are equal to one by construction. 

The results are presented in Tables~\ref{tab_result_single}
and~\ref{tab_result_electrons}, respectively, for the dimuon and dielectron
channels.
The two shape measurements, 
shown in the last column of the tables, 
are in good agreement for 11 out of 13 invariant
mass bins and remain statistically consistent (although marginally)
for the remaining two bins, 40--50\GeV and 120--150\GeV.

As a semi-independent check, a measurement was performed using a data
sample collected with a double-muon trigger with a lower \pt requirement of 7\GeV on each
muon. The signal yield is increased tenfold at the lowest invariant masses at the expense of
larger systematic uncertainties on the background. The result agrees with
the measurement made with the single muon trigger, having a similar precision in
the two lowest invariant mass bins.

\begin{table}[h!]
\begin{center}
\caption{Results for the DY spectrum normalized to the Z region in the
dimuon channel. The statistical and systematic uncertainties
are summed in quadrature.   $\RPOSTFSR$ and $\RPOSTFSRDET$ are
calculated using Eqs.~(\ref{eqn:fullCrossSectionRatio}) 
and~(\ref{eqn:fullCrossSectionRatio_DET}), respectively. The $\RDET$ and $R$ are
calculated using the FSR corrections given in Table~\ref{tab_accEff}.
\label{tab_result_single} }
\begin{tabular}{|l|r@{$~\pm~$}l|r@{$~\pm~$}l|r@{$~\pm~$}l|r@{$~\pm~$}l|}
\hline
Invariant mass bin (\!\GeV) 
& \multicolumn{2}{c|}{$\RPOSTFSRDET~(10^{-3})$} 
& \multicolumn{2}{c|}{$\RDET~(10^{-3}) $}
& \multicolumn{2}{c|}{$\RPOSTFSR~(10^{-3}) $}
& \multicolumn{2}{c|}{$R~(10^{-3}) $}\\
\hline
15--20   & 18 & 2 & 19 & 2 & 772 & 67 & 780 & 69 \\
20--30   & 58 & 3 & 58 & 3 & 528 & 33 & 533 & 34 \\
30--40   & 67 & 3 & 67 & 3 & 147 & 8  & 147 & 8 \\
40--50   & 44 & 2 & 41 & 2 & 66  & 4  & 62  & 4 \\
50--60   & 30 & 2 & 23 & 2 & 37  & 3  & 30  & 2 \\
60--76   & 51 & 2 & 28 & 1 & 55  & 3  & 32  & 2 \\
76--86   & 97 & 4 & 56 & 3 & 98  & 5  & 58  & 3 \\
86--96   & 803& 14& 861& 15& 799 & 23 & 857 & 26\\
96--106  & 38 & 3 & 43 & 3 & 37  & 3  & 41  & 3 \\
106--120 & 12 & 1 & 12 & 1 & 11  & 1  & 12  & 1 \\
120--150 & 9.2 & 0.9& 9.7 & 1.0 & 8.4 & 0.8 & 8.8 & 0.9\\
150--200 & 3.1 & 0.6& 3.2 & 0.7 & 2.6 & 0.5 & 2.7 & 0.6\\
200--600 & 1.8 & 0.4& 1.9 & 0.5 & 1.4 & 0.3 & 1.5 & 0.4\\
\hline
\end{tabular}
\end{center}
\end{table}

\begin{table}[h!]
\begin{center}
\caption{Results for the DY spectrum normalized to the Z region in the
dielectron channel. The statistical and systematic uncertainties
are summed in quadrature.   $\RPOSTFSR$ and $\RPOSTFSRDET$ are
calculated using Eqs.~(\ref{eqn:fullCrossSectionRatio}) and~(\ref{eqn:fullCrossSectionRatio_DET}), respectively. The $\RDET$ and $R$ are
calculated using the FSR corrections given in Table~\ref{tab_accEff_electrons}.
\label{tab_result_electrons} }
\begin{tabular}{|l|r@{$~\pm~$}l|r@{$~\pm~$}l|r@{$~\pm~$}l|r@{$~\pm~$}l|}
\hline
Invariant mass bin (\!\GeV) 
& \multicolumn{2}{c|}{$\RPOSTFSRDET~(10^{-3})$} 
& \multicolumn{2}{c|}{$\RDET~(10^{-3}) $}
& \multicolumn{2}{c|}{$\RPOSTFSR~(10^{-3}) $}
& \multicolumn{2}{c|}{$R~(10^{-3}) $}\\
\hline
15--20  & 6   & 3  & 6   & 3  & 487 & 230 & 508 & 238\\
20--30  & 13  & 2  & 13  & 2  & 536 & 96  & 559 & 97\\
30--40  & 24  & 4  & 22  & 4  & 129 & 22  & 131 & 21\\
40--50  & 28  & 4  & 24  & 4  & 52  & 8   & 47  & 7\\
50--60  & 30  & 5  & 19  & 3  & 39  & 6   & 27  & 4\\
60--76  & 78  & 12 & 30  & 4  & 84  & 13  & 36  & 5\\
76--86  & 144 & 60 & 61  & 25 & 147 & 60  & 64  & 26\\
86--96  & 722 & 62 & 839 & 60 & 715 & 62  & 834 & 60\\
96--106 & 44  & 21 & 55  & 26 & 43  & 20  & 53  & 25\\
106--120& 13  & 3  & 15  & 3  & 12  & 2   & 14  & 3\\
120--150& 5.4 & 1.2& 6.0 & 1.3& 4.8 & 1.1 & 5.4 & 1.2  \\
150--200& 2.5 & 0.8& 2.8 & 0.8& 2.1 & 0.6 & 2.3 & 0.7  \\
200--600& 2.1 & 0.6& 2.4 & 0.7& 1.5 & 0.5 & 1.7 & 0.5  \\ \hline
\end{tabular}
\end{center}
\end{table}

The theoretical cross section is calculated with FEWZ 
and three sets of PDFs: CT10, CTEQ66~\cite{CTEQ66}, and MSTW2008~\cite{MSTW2008}. 
The calculations include leptonic decays of Z bosons
with full spin correlations as well as nonzero width effects and
$\gamma^*$-Z interference.  However, they do not simulate FSR effects.
The calculations are cross-checked with the program DYNNLO, based on~\cite{DYNNLO,DYNNLO1},
which offers features similar to FEWZ. The predictions for the shape of
the DY spectrum agree well between the two programs, typically within 1\%. 

The uncertainties on the theoretical predictions due to the imprecise knowledge of the PDFs are calculated 
with the LHAGLUE 
interface to the PDF library 
LHAPDF~\cite{Bourilkov:2003kk,Whalley:2005nh}, using
a reweighting technique with asymmetric uncertainties~\cite{Bourilkov:2006cj}.
Since this is a shape measurement, and the normalization of the spectrum 
is defined by the number of events in the Z region, the uncertainty is
calculated for the yield ratio, $Y_i/Y_{\mathrm{norm}}$, where $Y_i$ is the 
predicted yield in the invariant mass bin~$i$ and $Y_{\mathrm{norm}}$ is the yield in the Z region.
The uncertainties for these ratios are much smaller than those for the individual yields 
because of the correlations between $Y_i$ and $Y_{\mathrm{norm}}$, especially in 
the dilepton invariant mass region close to the Z mass.

The factorization and renormalization scales were varied between 0.5 and 2 
times the dilepton invariant mass. The resulting variations of the cross sections at 
NNLO are much smaller than at NLO, and are less than 1.4\% around the Z peak.
The dependence of the DY cross section on the strong coupling constant 
$\alpha_s$ was evaluated by varying $\alpha_s$ between $0.116$ and $0.120$,
using FEWZ and the CT10 PDF set. The cross section variations are at the percent level.
Higher-order electroweak corrections for DY, evaluated with HORACE~\cite{HORACE}, 
showed a negligible influence (typically well below 1\%) on the shape measurements in 
the investigated invariant mass range.
The theoretical predictions from FEWZ at NNLO are presented in
Table~\ref{tab_theory_fullacc}.
 
\begin{table}[h!]
\begin{center}
\caption{Theoretical predictions at NNLO with FEWZ and three sets of PDFs.
The cross sections in this table are calculated in the full phase space with 
1\% statistical precision. The
theoretical predictions of the ratio $R$ and its uncertainties are also given. ``Other''
contains uncertainties from EWK correction, scale dependence, and $\alpha_s$.
\label{tab_theory_fullacc} }
\begin{tabular}{| l | l | l | l || l | l | l |}
\hline
Invariant mass & \multicolumn{3}{c||}{Cross section (pb)}& $R~(10^{-3})$  & \multicolumn{2}{c|}{Uncertainties on $R$ (\%)} \\
\cline{2-7}
bin (\!\GeV)&  CT10 & CTEQ66 & MSTW2008 &MSTW2008 & PDF & Other \\
\hline
15--20   & $787~~$ & $811~~$ & $819~~$ & $812~~$  & $+4.3$/$-3.3$ & $+2.5$/$-2.7$\\
20--30   & $476~~$ & $483~~$ & $499~~$ & $494~~$  & $+3.6$/$-2.8$ & $+1.9$/$-3.6$\\
30--40   & $135~~$ & $137~~$ & $142~~$ & $141~~$  & $+2.7$/$-2.3$ & $+3.1$/$-2.1$\\
40--50   & $53~~$ & $54~~$ & $56~~$ & $55~~$  & $+2.1$/$-1.9$ & $+2.4$/$-2.5$\\
50--60   & $27~~$ & $27~~$ & $29~~$ & $28~~$  & $+1.6$/$-1.5$ & $+2.6$/$-2.0$\\
60--76   & $32~~$ & $32~~$ & $33~~$ & $33~~$  & $+0.9$/$-0.9$ & $+2.0$/$-2.4$\\
76--86   & $56~~$ & $57~~$ & $58~~$ & $58~~$  & $+0.2$/$-0.2$ & $+2.1$/$-2.5$\\
86--96   & $822~~$ & $825~~$ & $852~~$ & $844~~$  & $+0.1$/$-0.1$ & $+1.8$/$-2.2$\\
96--106  & $51~~$ & $51~~$ & $53~~$ & $52~~$  & $+0.2$/$-0.2$ & $+2.8$/$-2.0$\\
106--120 & $12~~$ & $12~~$ & $13~~$ & $13~~$  & $+0.5$/$-0.5$ & $+2.6$/$-2.2$\\
120--150 & $6.7$ & $6.7$ & $7.0$ & $6.9$  & $+0.9$/$-0.9$ & $+2.5$/$-1.7$\\
150--200 & $2.6$ & $2.6$ & $2.7$ & $2.7$  & $+1.5$/$-1.6$ & $+2.0$/$-1.8$\\
200--600 & $1.3$& $1.3$ & $1.3$ & $1.3$  & $+2.8$/$-2.9$ & $+1.8$/$-2.1$\\
\hline
\end{tabular}
\end{center}
\end{table}

The results are also normalized to the invariant mass bin widths, $\Delta M_i$, defining
\begin{equation}\label{eqn:shape_r}
   r_i = \frac{R_i}{\Delta M_i} .
   \end{equation}

Assuming lepton universality, the dimuon and dielectron results for $r_i$ are
combined in a weighted average, using as weights the inverse of the
respective squared total uncertainties, where the statistical and
systematic uncertainties are added in quadrature.

The only expected source of correlation between the dimuon and dielectron results is due to the use
of the same MC model for the acceptance and FSR corrections.
Given that the uncertainties on these corrections are much smaller than most other uncertainties, 
especially in the dielectron channel, this correlation has a negligible influence on the 
combined results.

There are correlations between the invariant mass bins, induced by the various corrections 
applied in the analysis, especially those related to the efficiencies and resolutions.
The efficiency corrections are highly correlated between adjacent invariant mass bins, since
they tend to use the same T\&P factors, derived from the same single-lepton 
\pt bins. Nevertheless, for the dimuon channel the efficiency uncertainty is at most 20\%
of the total uncertainty, significantly diluting the effect of these correlations in the final results.
The resolution correlations, introduced through the unfolding procedure,
only have a visible effect around the Z peak. In summary, the level of correlations 
does not affect the combination of results in a significant way.

\begin{table}[h]
\begin{center}
\caption{Results for the DY spectrum normalized to the Z region and to the invariant mass bin width,
using Eq.~(\ref{eqn:shape_r}), before and after combining the two channels. The results presented are 
in {Ge\hspace{-.08em}V}$^{-1}$ units.
\label{tab_result_combined} }
\begin{tabular}{|l|r@{$~\pm~$}l|r@{$~\pm~$}l|r@{$~\pm~$}l|}
\hline
Invariant mass bin (\!\GeV) 
& \multicolumn{2}{c|}{$r$ (muons)}
& \multicolumn{2}{c|}{$r$ (electrons)}
& \multicolumn{2}{c|}{$r$ (combined)}\\
\hline
15--20    & $(15.6 $ & $1.4) \times 10^{-2}$  &  $(10.2 $ & $4.8) \times 10^{-2}$  &  $(15.2 $ & $1.3) \times 10^{-2}$ \\
20--30    & $(5.3 $ & $0.3) \times 10^{-2}$   &  $(5.6 $ & $1.0) \times 10^{-2}$   &  $(5.4 $ & $0.3) \times 10^{-2}$ \\
30--40    & $(1.5 $ & $ 0.1) \times 10^{-2}$  &  $(1.3 $ & $ 0.2) \times 10^{-2}$  &  $(1.5 $ & $ 0.1) \times 10^{-2}$ \\
40--50    & $(6.2 $ & $ 0.4) \times 10^{-3}$  &  $(4.7 $ & $ 0.7) \times 10^{-3}$  &  $(5.9 $ & $ 0.3) \times 10^{-3}$ \\
50--60    & $(3.0 $ & $ 0.2) \times 10^{-3}$  &  $(2.7 $ & $ 0.4) \times 10^{-3}$  &  $(3.0 $ & $ 0.2) \times 10^{-3}$ \\
60--76    & $(2.0 $ & $ 0.1) \times 10^{-3}$  &  $(2.2 $ & $ 0.3) \times 10^{-3}$  &  $(2.1 $ & $ 0.1) \times 10^{-3}$ \\
76--86    & $(5.8 $ & $ 0.3) \times 10^{-3}$  &  $(6.4 $ & $ 2.6) \times 10^{-3}$  &  $(5.8 $ & $ 0.3) \times 10^{-3}$ \\
86--96    & $(85.7 $ & $ 2.6) \times 10^{-3}$ &  $(83.4 $ & $ 6.0) \times 10^{-3}$ &  $(85.6 $ & $ 2.4) \times 10^{-3}$ \\
96--106   & $(4.1 $ & $ 0.3) \times 10^{-3}$  &  $(5.3 $ & $ 2.5) \times 10^{-3}$  &  $(4.2 $ & $ 0.3) \times 10^{-3}$ \\
106--120  & $(8.4 $ & $ 0.9) \times 10^{-4}$  &  $(9.6 $ & $ 1.9) \times 10^{-4}$  &  $(8.6 $ & $ 0.8) \times 10^{-4}$ \\
120--150  & $(2.9 $ & $ 0.3) \times 10^{-4}$  &  $(1.8 $ & $ 0.4) \times 10^{-4}$  &  $(2.5 $ & $ 0.2) \times 10^{-4}$ \\
150--200  & $(5.4 $ & $ 1.2) \times 10^{-5}$  &  $(4.6 $ & $ 1.4) \times 10^{-5}$  &  $(5.1 $ & $ 0.9) \times 10^{-5}$ \\
200--600  & $(3.7 $ & $ 1.0) \times 10^{-6}$  &  $(4.3 $ & $ 1.3) \times 10^{-6}$  &  $(3.9 $ & $ 0.8) \times 10^{-6}$ \\
\hline
\end{tabular}
\end{center}
\end{table}

\begin{figure}[h!]
{\centering
\includegraphics[width=0.80\textwidth, angle=90]{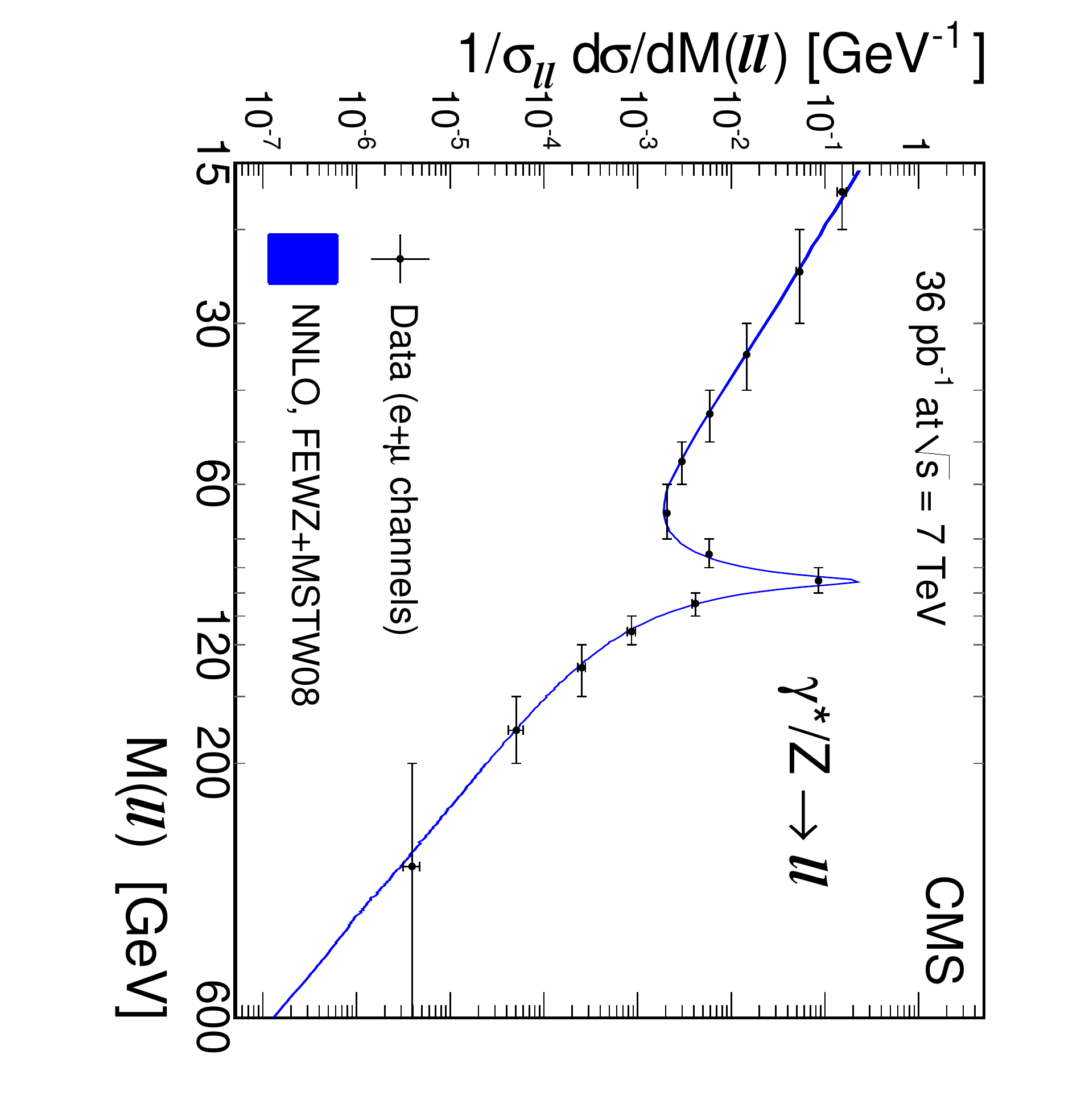}
\caption{\label{results}
DY invariant mass spectrum, normalized to the Z resonance region, 
$r = (1/\sigma_{\mathrm{\ell\ell}}) d\sigma/d\MLL$, as measured 
and as predicted by
NNLO calculations, for the full phase space. The vertical error bar indicates 
the experimental (statistical and systematic) uncertainties summed in quadrature with
the theory uncertainty
resulting from the model-dependent kinematic distributions inside each bin.
The horizontal bars indicate bin sizes and the data points inside are placed according to  Ref.~\cite{binCorrection}. The width of the theory curve represents uncertainties from Table \ref{tab_theory_fullacc}.}}
\end{figure}

Table~\ref{tab_result_combined} gives the measured shape $r$, defined in Eq.~(\ref{eqn:shape_r}), 
both in the dimuon and dielectron channels and also the combined result.
Figure~\ref{results} compares the measured (combined)
results with the prediction from the FEWZ NNLO calculations, performed with the MSTW08 PDF set. To provide
a meaningful comparison, 
each data point is located on the horizontal axis at the position where 
the theoretical function has a value equal to its mean value over the 
corresponding bin, following
the procedure described
in Ref.~\cite{binCorrection}. The measurements are very well reproduced by the theoretical calculations.

\section{Summary}
\label{sec:conclusions}

The Drell--Yan differential cross section normalized to the cross section in the Z region
has been measured in $\pp$ collisions at $\sqrt{s} = 7\TeV$,
in the dimuon and dielectron channels in the invariant mass range $15 < \MLL < 600 \GeV$.
The measurement is based on event samples collected by the CMS experiment,
corresponding to an integrated luminosity of $35.9\pm 1.4~{\mathrm{pb}}^{-1}$.
Results are presented both inside the detector acceptance and in the full phase space,
and the effect of final state QED radiation on the results is reported as well.
A correct description of the measurements requires modeling to NNLO
for dilepton invariant masses below about 30 GeV.
The measurements are in good agreement with the NNLO theoretical predictions,
as computed with {\sc FEWZ}.

\section*{Acknowledgments}

\hyphenation{Bundes-ministerium Forschungs-gemeinschaft Forschungs-zentren} 
We would like to thank the authors of {\sc FEWZ} and {\sc POWHEG} for the fruitful discussions, co-operation, and cross-checks in performing the theoretical calculations for our analysis.

We wish to congratulate our colleagues in the CERN accelerator departments for the excellent performance of the LHC machine. We thank the technical and administrative staff at CERN and other CMS institutes. This work was supported by the Austrian Federal Ministry of Science and Research; the Belgium Fonds de la Recherche Scientifique, and Fonds voor Wetenschappelijk Onderzoek; the Brazilian Funding Agencies (CNPq, CAPES, FAPERJ, and FAPESP); the Bulgarian Ministry of Education and Science; CERN; the Chinese Academy of Sciences, Ministry of Science and Technology, and National Natural Science Foundation of China; the Colombian Funding Agency (COLCIENCIAS); the Croatian Ministry of Science, Education and Sport; the Research Promotion Foundation, Cyprus; the Estonian Academy of Sciences and NICPB; the Academy of Finland, Finnish Ministry of Education and Culture, and Helsinki Institute of Physics; the Institut National de Physique Nucl\'eaire et de Physique des Particules~/~CNRS, and Commissariat \`a l'\'Energie Atomique et aux \'Energies Alternatives~/~CEA, France; the Bundesministerium f\"ur Bildung und Forschung, Deutsche Forschungsgemeinschaft, and Helmholtz-Gemeinschaft Deutscher Forschungszentren, Germany; the General Secretariat for Research and Technology, Greece; the National Scientific Research Foundation, and National Office for Research and Technology, Hungary; the Department of Atomic Energy and the Department of Science and Technology, India; the Institute for Studies in Theoretical Physics and Mathematics, Iran; the Science Foundation, Ireland; the Istituto Nazionale di Fisica Nucleare, Italy; the Korean Ministry of Education, Science and Technology and the World Class University program of NRF, Korea; the Lithuanian Academy of Sciences; the Mexican Funding Agencies (CINVESTAV, CONACYT, SEP, and UASLP-FAI); the Ministry of Science and Innovation, New Zealand; the Pakistan Atomic Energy Commission; the State Commission for Scientific Research, Poland; the Funda\c{c}\~ao para a Ci\^encia e a Tecnologia, Portugal; JINR (Armenia, Belarus, Georgia, Ukraine, Uzbekistan); the Ministry of Science and Technologies of the Russian Federation, the Russian Ministry of Atomic Energy and the Russian Foundation for Basic Research; the Ministry of Science and Technological Development of Serbia; the Ministerio de Ciencia e Innovaci\'on, and Programa Consolider-Ingenio 2010, Spain; the Swiss Funding Agencies (ETH Board, ETH Zurich, PSI, SNF, UniZH, Canton Zurich, and SER); the National Science Council, Taipei; the Scientific and Technical Research Council of Turkey, and Turkish Atomic Energy Authority; the Science and Technology Facilities Council, UK; the US Department of Energy, and the US National Science Foundation.

Individuals have received support from the Marie-Curie programme and the European Research Council (European Union); the Leventis Foundation; the A. P. Sloan Foundation; the Alexander von Humboldt Foundation; the Associazione per lo Sviluppo Scientifico e Tecnologico del Piemonte (Italy); the Belgian Federal Science Policy Office; the Fonds pour la Formation \`a la Recherche dans l'Industrie et dans l'Agriculture (FRIA-Belgium); the Agentschap voor Innovatie door Wetenschap en Technologie (IWT-Belgium); and the Council of Science and Industrial Research, India; the European Union Structural Funds project `Postdoctoral Fellowship Implementation in Lithuania'.

\bibliography{auto_generated}   

\cleardoublepage \appendix\section{The CMS Collaboration \label{app:collab}}\begin{sloppypar}\hyphenpenalty=5000\widowpenalty=500\clubpenalty=5000\textbf{Yerevan Physics Institute,  Yerevan,  Armenia}\\*[0pt]
S.~Chatrchyan, V.~Khachatryan, A.M.~Sirunyan, A.~Tumasyan
\vskip\cmsinstskip
\textbf{Institut f\"{u}r Hochenergiephysik der OeAW,  Wien,  Austria}\\*[0pt]
W.~Adam, T.~Bergauer, M.~Dragicevic, J.~Er\"{o}, C.~Fabjan, M.~Friedl, R.~Fr\"{u}hwirth, V.M.~Ghete, J.~Hammer\cmsAuthorMark{1}, S.~H\"{a}nsel, M.~Hoch, N.~H\"{o}rmann, J.~Hrubec, M.~Jeitler, W.~Kiesenhofer, M.~Krammer, D.~Liko, I.~Mikulec, M.~Pernicka, B.~Rahbaran, H.~Rohringer, R.~Sch\"{o}fbeck, J.~Strauss, A.~Taurok, F.~Teischinger, C.~Trauner, P.~Wagner, W.~Waltenberger, G.~Walzel, E.~Widl, C.-E.~Wulz
% ...
\vskip\cmsinstskip
% ...
1:~~Also at CERN, European Organization for Nuclear Research, Geneva, Switzerland\\
% ...

\end{sloppypar}
\end{document}